\documentclass[prb,superscriptaddress]{revtex4}

\usepackage{epsfig}
\usepackage{bm}

\begin{document}
\title{Random Kondo alloys }

\author{S. Burdin}
\affiliation{
Max-Planck-Institut f\"{u}r Physik komplexer Systeme, 
N\"{o}thnitzer Strasse 38, 01187 Dresden, Germany}

\author{P. Fulde}
\affiliation{
Max-Planck-Institut f\"{u}r Physik komplexer Systeme, 
N\"{o}thnitzer Strasse 38, 01187 Dresden, Germany}

\maketitle
%
\begin{center}
{\bf Abstract }
\end{center}
The interplay between the Kondo effect and disorder is studied. This is done
by applying a matrix coherent potential approximation (CPA) and treating the
Kondo interaction on a mean-field level. The resulting equations are shown to
agree with those derived by the dynamical mean-field method (DMFT). By
applying the formalism to a Bethe tree structure with infinite coordination
the effect of diagonal and off-diagonal disorder are studied. Special
attention is paid to the behavior of the Kondo- and the Fermi liquid
temperature as function of disorder and concentration of the Kondo ions. The
non monotonous dependence of these quantities is discussed. 


\section{Introduction}
The Kondo effect is one of the most investigated phenomena in solid-state 
physics. Part of the reason is that it can not be treated perturbationally 
since it is a strong coupling effect. Therefore it requires special theoretical
tools to deal with it. Kondo physics occurs when strongly correlated electrons
like $4f$ electrons in $Ce^{3+}$ or holes in $Yb^{3+}$ are weakly hybridizing 
with the
conduction electrons of their surrondings. This results in low-energy excitations
which in the case of concentrated systems may result in heavy
quasiparticles. For recent reviews of the field we refer to
Refs.~[\onlinecite{reviewHF1}-\onlinecite{reviewHF2}-\onlinecite{KondoHewson}]. 
A realistic starting point for Kondo systems is the Anderson impurity 
or Anderson lattice model. Due to the hybridization mentioned above it involves
spin as well as charge degrees of freedom. 
Often the charge degrees of freedom are less interesting and are therefore
eliminated by a Schrieffer-Wolf transformation~[\onlinecite{SchriefferWolf}]. 
The result is an
antiferromagnetic interaction between the spins of the conduction electrons
and the strongly correlated localised, e.g., $4f$ electrons. This leads to the Kondo
Hamiltonian. 

Competing with the Kondo effect is the Ruderman-Kittel-Kasuya-Yoshida 
(RKKY) interaction. While the Kondo effect leads to the formation of a singlet
between the spins of the $4f$ and conduction electrons, the RKKY interaction lowers the
energy of a system of local spins interacting with each other via conduction
electrons. Therefore if the latter is more important than the former, the local
spins will remain uncompensated and eventually order and not participate in the singlet
formation. 

The aim of the present investigation is to study the effect of disorder on
Kondo
physics~[\onlinecite{Kondoanddisorder1}-\onlinecite{Kondoanddisorder2}-\onlinecite{Kondoanddisorder2bis}-\onlinecite{Kondoanddisorder3}-\onlinecite{Kondoanddisorder4}-\onlinecite{Kondoanddisorder5}-\onlinecite{Vojta}]. 
It has been suggested in several works that disorder leads to 
non-Fermi-liquid behavior (NFL) at low temperatures. 
For example, it has been shown in 
Refs.~[\onlinecite{Kondoanddisorder1}-\onlinecite{Kondoanddisorder2}-\onlinecite{Kondoanddisorder5}] that 
a distribution of Kondo temperatures $T_{K}$ can result from local disorder, 
the NFL features beeing related to the presence of
very-low-$T_{K}$ spins which remain unquenched at any finite temperature. 
Another possible scenario attributes the NFL behavior to the proximity to a
quantum critical point resulting from disordered RKKY 
interactions~[\onlinecite{Kondoanddisorder2bis}-\onlinecite{Kondoanddisorder3}-\onlinecite{Kondoanddisorder4}]. 
More recently, it has been suggested that a NFL behavior can occur between the 
local Fermi-liquid (FL) and coherent heavy FL phases characterizing
respectively a diluted and a dense Kondo alloy~[\onlinecite{Vojta}]. 

We assume that we are in a regime where the Kondo effect is
more important than the RKKY interaction so that the latter may be neglected. 
Instead we concentrate on the singlet formation energy and on how it can be 
expressed in terms of the Kondo temperature $T_{K}$ and of the temperature 
$T_{FL}$ at which the different low energy excitations form coherent
quasiparticles. In particular we study how $T_K$ and $T_{FL}$ behave as 
functions of conduction electron band filling $n_{c}$, 
local spin concentration $x$ and disorder.

\section{Model Hamiltonian and methods of solutions}

We consider the Kondo alloy model (KAM) with the Hamiltonian 
\begin{eqnarray} 
H =\sum_{ij\sigma}t_{ij}c_{i\sigma}^{\dagger}c_{j\sigma}
+
\frac{J_K}{2}\sum_{i\in A}\sum_{\sigma\sigma '}
S_{i}^{\sigma\sigma '}c_{i\sigma '}^{\dagger}c_{i\sigma}~, 
\label{HamiltonianKAM}
\end{eqnarray}
where the first term describes nearest-neighbor hopping of conduction electrons 
on a lattice with sites occupied randomly by atoms of kind $A$ and $B$. 
The corresponding concentrations are $c_A=x$ and $c_B=1-x$. The hopping matrix 
elements have three different values, i.e., 
\begin{eqnarray}
t_{ij}=\gamma_{ij} \left[
\begin{array}{lll}
t_{A} & $if$ & i,j\in A\\
t_{B} & $if$ & i,j\in B\\
t_{AB}& $otherwise$ & \\
\end{array}
\right .~. 
\label{Definitionhopping}
\end{eqnarray}
Here $\gamma_{ij}$ is the structure factor of the underlying periodic 
lattice with Fourier transform $\gamma_{{\bf k}}\equiv
\sum_{ij}\gamma_{ij}\exp{i{\bf k}.({\bf R}_i-{\bf R}_{j})}$.  
The second term in Eq.~(\ref{HamiltonianKAM}) describes 
the Kondo interaction between the conduction electron spin and local 
spin operators ${\bf S}_{i}$, 
the latter being attached to atoms of type $A$ only. 

We shall treat the Kondo alloy model defined by Eq.~(\ref{HamiltonianKAM}) by 
applying a number of approximations. A rather simple one is that we assume a
random distribution of sites $A$ and $B$. As regards the Kondo interaction 
we shall consider two different ways of treating the randomness. They are
similar to each other but based on different physical pictures. 

One approach is a generalisation of a CPA matrix approach originally
introduced in Refs~[\onlinecite{matrixCPABEB1}-\onlinecite{matrixCPABEB2}]. 
The Kondo interaction is treated here within a mean-field approximation. 
The second approach is a matrix generalisation of the dynamical mean field
theory (DMFT) which is exact in the
limit of infinite dimensions~[\onlinecite{reviewDMFT1}-\onlinecite{reviewDMFT2}]. 
Averaging over the randomness is done here without simplifying the Kondo
interaction. A mean-field approximation can be introduced before or after the
DMFT approximation and leads to the same set of self-consistent equations as
obtained in the first, i.e., the generalised CPA approach. 

The analytical expressions obtained from these two approaches are
applicable to any lattice structure. 
The numerical results presented below apply the DMFT to a Bethe lattice 
instead of a regular one. 
The Kondo interaction is treated in this case within the mean-field approximation. 

\section{The matrix-CPA method}

\subsection{Mean-field treatment of the Kondo interaction}
We begin with a mean-field approximation for the Kondo interaction. 
Following the standard
theory~[\onlinecite{Kondomeanfield1}-\onlinecite{Kondomeanfield2}-\onlinecite{KondolargeN1}-\onlinecite{KondolargeN5}-\onlinecite{KondolargeN6}], 
the spin 
operators are written in the fermionic representation 
${\bf S}_{i}^{\sigma\sigma '}=
f_{i\sigma}^{\dagger}f_{i\sigma '}-\delta_{\sigma\sigma '}/2$, 
with the constraint 
$\sum_{\sigma}f_{i\sigma}^{\dagger}f_{i\sigma}=1$. 
The Hamiltonian Eq.~(\ref{HamiltonianKAM}) becomes therefore
\begin{eqnarray} 
H = \sum_{ij\sigma}t_{ij}c_{i\sigma}^{\dagger}c_{j\sigma}
+
\frac{J_K}{2}\sum_{i\in A}\sum_{\sigma\sigma '}
f_{i\sigma}^{\dagger}f_{i\sigma '}c_{i\sigma '}^{\dagger}c_{i\sigma}~. 
\label{HamiltonianKAMSUN}
\end{eqnarray}
The systems we want to describe here involve physical spins $1/2$ 
with a $SU(2)$ symmetry. 
The mean-field approach as 
introduced in Refs~[\onlinecite{Kondomeanfield1}-\onlinecite{Kondomeanfield2}] is in this case an
approximation which 
becomes exact in the limit of $SU(N\to \infty)$ 
symmetry~[\onlinecite{KondolargeN1}-\onlinecite{KondolargeN5}-\onlinecite{KondolargeN6}]. 
The Hamiltonian is
\begin{eqnarray} 
H =\sum_{ij\sigma}t_{ij}c_{i\sigma}^{\dagger}c_{j\sigma}
+
r
\sum_{\sigma}
\sum_{i\in A}
\left (
c_{i\sigma}^{\dagger}f_{i\sigma}+f_{i\sigma}^{\dagger}c_{i\sigma}
\right )
-
\mu
\sum_{\sigma}
\sum_{i}
\left( c_{i\sigma}^{\dagger}c_{i\sigma}-{n_{c}\over{2}}\right)
-
\lambda
\sum_{\sigma}
\sum_{i\in A}
\left( f_{i\sigma}^{\dagger}f_{i\sigma}-{1\over{2}}\right)~. 
\label{HamiltonianlargeN}
\end{eqnarray}
where $n_{c}$ is the average number of conduction electrons per site 
$i$ while $\mu$ denotes the chemical potential. 
In the following, we discard the spin index $\sigma$ since 
in mean-field approximation the contributions to $H$ of the different spin 
components
decouple~[\onlinecite{Kondomeanfield1}-\onlinecite{Kondomeanfield2}-\onlinecite{KondolargeN1}-\onlinecite{KondolargeN5}-\onlinecite{KondolargeN6}]. 
The Kondo interaction is approximated by an effective hybridization 
 $r=J_K [\langle f_{i}c_{i}^{\dagger}\rangle ]$ 
between the conduction electrons and the fermionic operators, 
where the $\langle\cdots\rangle$ denotes the thermal average with respect 
to the Hamiltonian (\ref{HamiltonianlargeN}) for random 
configurations of sites $A$ and $B$, and $[\cdots ]$ denotes the average 
with respect to these configurations. 
Note that the same form of the Hamiltonian is obtained by starting from 
an Anderson lattice instead of a Kondo lattice, and treating it within the 
mean-field slave boson approximation~[\onlinecite{KondolargeN1}-\onlinecite{KondolargeN5}-\onlinecite{KondolargeN6}]. 
We have started here from the Kondo Hamiltonian because we are interested in
the case of near integer valency  of the impurity, i.e., the $f$ electron 
count is supposed to be very close to one. 
The above mean-field approximation leads to an $f-$like band. It
models the low-energy excitations which result from the Kondo interaction 
or alternatively Anderson Hamiltonian. 
The conditions $\sum_{\sigma}f_{i\sigma}^{\dagger}f_{i\sigma}=1$ are taken
into account by Lagrange parameters $\lambda_{i}$. We set all of them equal to 
$\lambda$ which implies that the above conditions are satisfied on average only. 
Thus small local fluctuations in the $f$ electron count are possible here 
like in the Anderson model. 

The quantities $\mu$,
$\lambda$ and $r$ are determined by self-consistency conditions. For that
purpose local Green's functions are introduced. They are different for
magnetic sites $A$, nonmagnetic sites $B$ and for $f$ as well as conduction 
electrons. The $G_{ij}^{ff}(\tau-\tau' )\equiv 
-\langle T_{\tau} f_{i}(\tau)f_{j}^{\dagger}(\tau')\rangle$,  
$G_{ij}^{fc}(\tau-\tau ')\equiv
-\langle T_{\tau} f_{i}(\tau)c_{j}^{\dagger}(\tau')\rangle$, 
and 
$G_{ij}^{cc}(\tau-\tau' )\equiv 
-\langle T_{\tau} c_{i}(\tau)c_{j}^{\dagger}(\tau')\rangle$ are finite
temperature Green's functions defined for imaginary time $\tau$, where 
$T_{\tau}$ denotes 
the imaginary-time chronological ordering. 
We also define the averaged local Green's functions 
\begin{eqnarray}
G_{A}^{ff}\equiv {1\over{x}}\sum_{i\in A}G_{ii}^{ff}~, 
&& G_{A}^{fc}\equiv {1\over{x}}\sum_{i\in A}G_{ii}^{fc}~, \\
G_{A}^{cc}\equiv {1\over{x}}\sum_{i\in A}G_{ii}^{cc}~, 
&& G_{B}^{cc}\equiv {1\over{1-x}}\sum_{i\in B}G_{ii}^{cc}~. 
\end{eqnarray}

The chemical potential $\mu$, the Lagrange multiplier $\lambda$, and the
effective hybridization $r$ are determined by the self-consistent saddle point
equations: 
\begin{eqnarray}
-r/J_K
=&  {1\over{x}}\sum_{i\in A}\langle f_{i}c_{i}^{\dagger} \rangle
&=
G^{fc}_{A}(\tau=0^-)~, 
\label{EquationlargeN1}
\\
1/2
=& 
{1\over{x}}\sum_{i\in A}\langle f_{i}^{\dagger} f_{i}\rangle
&=
G^{ff}_{A}(\tau=0^-)~, 
\label{EquationlargeN2}
\\
n_c/2
=& 
\sum_{i}\langle c_{i}^{\dagger} c_{i}\rangle
&=
G^{cc}(\tau=0^-)~, 
\label{EquationlargeN3}
\end{eqnarray}
with $G^{cc}\equiv \sum_{i}G_{ii}^{cc} =xG_{A}^{cc}+(1-x)G_{B}^{cc}$. 

\subsection{Configuration averages}
For the determination of 
the Green's function of the conduction electrons we choose a generalization of 
the Coherent Potential Approximation (CPA) to a matrix form as 
introduced in Refs~[\onlinecite{matrixCPABEB1}-\onlinecite{matrixCPABEB2}]. 
Within that approximation, the system can be viewed as a medium 
with three interacting fermionic bands: 
two bands corresponding to conduction electrons on sites $A$ or $B$, 
and a third one representing the excitations of the strongly correlated 
$f-$electrons. 
Therefore the dynamics related to the spins of the $A$ sites is described in a
simplified form, i.e., in the form of $f-$electrons with a dispersive band. 
The $3\times 3$ Green's function matrix is of the following form 
\begin{eqnarray}
{\bf\widetilde{G}}_{ij}=  
\left(
\begin{array}{ccc}
\hat{x}_{i}\hat{x}_{j}G_{ij}^{ff} & \hat{x}_{i}\hat{x}_{j}G_{ij}^{fc}& \hat{x}_{i}\hat{y}_{j}G_{ij}^{fc} \\
\hat{x}_{i}\hat{x}_{j}G_{ij}^{cf} & \hat{x}_{i}\hat{x}_{j}G_{ij}^{cc}& \hat{x}_{i}\hat{y}_{j}G_{ij}^{cc} \\
\hat{y}_{i}\hat{x}_{j}G_{ij}^{cf} & \hat{y}_{i}\hat{x}_{j}G_{ij}^{cc}& \hat{y}_{i}\hat{y}_{j}G_{ij}^{cc} 
\end{array}
\right)~,
\label{ExpressionGreemmatrixCPA}
\end{eqnarray}
where $\hat{x}_{i}=1-\hat{y}_{i}$ are projection operators, which are unity (zero) if 
site $i$ is occupied by an $A$ ($B$) atom. 
Averaging the local Green's function matrix with respect to 
different configurations of randomly
distributed types of atoms we find
\begin{eqnarray}
&\left[ {
{\bf \widetilde{G}}_{ii}
}\right]
&=
\left (
\begin{array}{ccc}
xG^{ff}_{A} & xG^{fc}_{A} & 0 \\
xG^{cf}_{A} & xG^{cc}_{A} & 0 \\
0          & 0          & (1-x)G^{cc}_{B}
\end{array}
\right )~. 
\end{eqnarray}
In this expression, the vanishing of the mixed $A-B$ matrix elements 
follows 
directly from $\hat{x}_{i}\hat{y}_{i}=0$, which ensures that a given site 
is either 
of kind $A$ or $B$. 
Averaging over the different configurations of $A$ and $B$ sites restores lattice 
translation symmetry. Therefore we define
\begin{eqnarray}
{\bf \widetilde{G}}_{{\bf k}}
\equiv 
\sum_{ij}e^{-i{\bf k}.({\bf R}_{i}-{\bf R}_{j})}
\left[ { {\bf \widetilde{G}}_{ij} }\right]
\label{DefinitionGreenreciprocalspaceCPA}
\end{eqnarray}
Within the single component CPA, the system is approximated by an 
effective medium, 
characterised by a local, i.e., ${\bf k}-$independent, but frequency dependent
self-energy~[\onlinecite{scalarCPA1}-\onlinecite{scalarCPA2}-\onlinecite{scalarCPA3}-\onlinecite{scalarCPA4}-\onlinecite{scalarCPA5}]. 
The latter is determined
self-consistently by requiring that the scattering matrix of the atoms $A$ and
$B$ within this effective medium vanishes on average. 
The matrix form of the CPA introduced in 
Refs~[\onlinecite{matrixCPABEB1}-\onlinecite{matrixCPABEB2}]
generalises the scalar procedure to an effective medium with two bands of
conduction electrons. 
Here we generalise the $2\times 2$ matrix form of the CPA to a $3\times 3$ one. 
The averaged Green's function matrix characterising the effective medium 
is given by the relation
\begin{eqnarray}
\left(
\left[
{\bf \widetilde{G}}(i\omega_{n})
\right]^{-1}
\right)_{ij}
=
i\omega_{n}{\bf \widetilde{I}}\delta_{ij}
-{\bf \widetilde{\Sigma}}(i\omega_{n})\delta_{ij}
-{\bf \widetilde{W}}\gamma_{ij}~,  
\label{RelationGreenrealspaceCPA}
\end{eqnarray}
where $i\omega_{n}\equiv i\pi T(2n +1)$ denotes the fermionic Matsubara
frequencies. In the following we leave out the $n$ index. 
Invoking the reciprocal space Green's function matrix defined by 
Eq.~(\ref{DefinitionGreenreciprocalspaceCPA}), the 
relation~(\ref{RelationGreenrealspaceCPA}) becomes 
\begin{eqnarray}
{\bf \widetilde{G}}_{{\bf k}}^{-1}(i\omega)
=
i\omega{\bf \widetilde{I}}
-
{\bf \widetilde{\Sigma}}(i\omega)
-
{\bf \widetilde{W}}\gamma_{{\bf k}}~. 
\label{ExpressionGreenreciprocalCPA}
\end{eqnarray}
Here ${\bf \widetilde{I}}$ is a $3\times 3$ unit matrix, 
${\bf \widetilde{W}}$ is the transfer matrix, 
\begin{eqnarray}
{\bf \widetilde{W}}=
\left(
\begin{array}{lll}
0 & 0 & 0 \\
0 & t_{A} & t_{AB} \\
0 & t_{AB}& t_{B} 
\end{array}
\right)~,  
\end{eqnarray}
and ${\bf \widetilde{\Sigma}}$ is a local self-energy matrix, 
\begin{eqnarray}
{\bf \widetilde{\Sigma}}(i\omega)\equiv
\left(
\begin{array}{cc}
{\bf \Sigma}_{A} (i\omega)
& 
\begin{array}{c}
\sigma_{1} (i\omega)\\ 
\sigma_{2}(i\omega)
\end{array} \\
\begin{array}{cc}
\sigma_{1}(i\omega) & \sigma_{2}(i\omega)
\end{array} 
&
\Sigma_{B}(i\omega)
\end{array}
\right)~. 
\label{DefinitionSigmaCPA}
\end{eqnarray}
which is determined by the set of self-consistent equations 
(see appendix \ref{AppendixCPAequations}): 
\begin{eqnarray}
&\left[
{\bf \widetilde{G}}_{ii}(i\omega)
\right]
&=
\left (
\begin{array}{ccc}
xG^{ff}_{A}(i\omega) & xG^{fc}_{A}(i\omega) & 0 \\
xG^{cf}_{A}(i\omega) & xG^{cc}_{A}(i\omega) & 0 \\
0          & 0          & (1-x)G^{cc}_{B}(i\omega)
\end{array}
\right )
=\sum_{{\bf k}}{\bf \widetilde{G}}_{{\bf k}}(i\omega)~, 
\label{RelationGreenreciprocalCPA}
\\
&&\nonumber\\
&{\bf \Sigma}_{A}(i\omega)
&=
-
\left ( {
\begin{array}{lr}
\lambda & r \\
r & \mu 
\end{array}
}\right )
-{{(1-x)}\over{x}}
\left ( {
\begin{array}{cc}
G_{A}^{ff}(i\omega) & G_{A}^{fc}(i\omega) \\
G_{A}^{cf}(i\omega) & G_{A}^{cc}(i\omega) 
\end{array}
}\right )^{-1}~, 
\label{RelationSigmaACPA}
\\
&&\nonumber\\
&\Sigma_{B}(i\omega)
&=
-
\mu 
-{x\over{(1-x)G_{B}^{cc}(i\omega)}}~. 
\label{RelationSigmaBCPA}
\end{eqnarray}
The self-energies $\sigma_{1}$ and $\sigma_{2}$ are determined 
by requiring that the mixed $A-B$ elements of the 
local Green's function matrix in Eq.~(\ref{RelationGreenreciprocalCPA}) 
vanish for the same reason as in Eq.~(\ref{ExpressionGreemmatrixCPA}). 
We find that 
\begin{eqnarray}
\sigma_{1}(i\omega)=0~, 
\end{eqnarray}
which reflects the fact that there is no direct interaction between 
$f-$fermions and 
the $B-$electronic band, describing the electrons on nonmagnetic sites. 
We find also an explicit expression for $\sigma_{2}$, i.e., 
\begin{eqnarray}
\sigma_{2}(i\omega)
=
-t_{AB}
{
{1-2x
+x(i\omega+\mu-r^2/(i\omega+\lambda))G_{A}^{cc}(i\omega)
-(1-x)(i\omega+\mu)G_{B}^{cc}(i\omega)
}
\over{
xt_{A}G_{A}^{cc}(i\omega)
-(1-x)t_{B}G_{B}^{cc}(i\omega)
}}~, 
\label{Expressionsigma2CPA}
\end{eqnarray}
which results from the direct hopping of conduction electrons 
between $A$ and $B$ sites. 

A complete solution of the initial Kondo alloy system is obtained by 
solving simultaneously the mean-field 
Eqs.~(\ref{EquationlargeN1}, 
\ref{EquationlargeN2}, 
\ref{EquationlargeN3}) 
together with the matrix Eq.~(\ref{RelationGreenreciprocalCPA}). Thereby the 
${\bf k}-$dependent average Green's function matrix is determined by the 
relation Eq.~(\ref{ExpressionGreenreciprocalCPA}), with the lattice
structure factor $\gamma_{{\bf k}}$ and the local self-energy matrix 
${\widetilde{\bf \Sigma}}$. The latter is determined by the self-consistent CPA 
Eqs.~(\ref{DefinitionSigmaCPA},  
\ref{RelationSigmaACPA}, 
\ref{RelationSigmaBCPA}).

\subsection{Kondo temperature}
We define the Kondo temperature $T_{K}$ as the temperature at which  
the effective hybridization $r$ (obtained from 
Eqs.~(\ref{EquationlargeN1}, 
\ref{EquationlargeN2}, 
\ref{EquationlargeN3})) vanishes. 
We find 
\begin{eqnarray}
2/J_{K}=\int
d\epsilon 
\rho_{A}^{0}(\epsilon+\mu_{0})
tanh[(\epsilon/2T_{K}]/\epsilon~, 
\label{EquationTK}
\end{eqnarray}
where $\rho_{A}^{0}$ and $\mu_{0}$ are respectively the local electronic 
density of states (DOS) on a magnetic site, and the chemical potential 
of a random $A-B$ alloy 
without Kondo interaction. 
An explicit expression for $T_{K}$ has been derived in 
Ref.~[\onlinecite{KondolatticelargeN}] in the weak-coupling regime 
$J_{K}<<t_{A}, t_{B}, t_{AB}$: 
\begin{eqnarray}
T_{K}=De^{-1/(J_{K}\rho_{A}^{0}(E_{F}))}\sqrt{1-(E_{F}/D)^{2}}F_{K}(n_{c})~, 
\label{ExplicitTK}
\end{eqnarray}
\begin{eqnarray}
F_{K}(n_{c})=\exp{\left( {
\int_{-(D+E_{F})}^{D-E_{F}}
\frac{d\omega}{\vert \omega\vert }
\frac{\rho_{A}^{0}(E_{F}+\omega )-\rho_{A}^{0}(E_{F})}{2\rho_{A}^{0}(E_{F})}
}\right) }~, 
\end{eqnarray}
where $D$ is the half-bandwidth of the non-interacting local DOS 
$\rho_{A}^{0}(\omega )$.

\section{DMFT equations}
The matrix form of the CPA introduced in Refs.~[\onlinecite{matrixCPABEB1}-\onlinecite{matrixCPABEB2}] 
was generalized to the KAM~(\ref{HamiltonianKAM}) after an appropriate 
mean-field approximation was made for the Kondo interaction. It allowed for
keeping the dynamical aspects of the $A$ sites with their attached spins by
means of introducing an additional $f-$like band of excitations. It 
supplemented the two bands resulting from the conduction electrons of the $A$
and $B$ sites. 
In the following we develop for the KAM a matrix DMFT computational scheme 
which can be formulated without a mean-field approximation for the Kondo
interaction. Note that our approach is different from the dynamical cluster
approximation introduced in Ref.~[\onlinecite{ClusterDMFTCPA}] since the
latter concerns diagonal disorder only. 

\subsection{DMFT matrix formalism for a binary alloy}
The Kondo-alloy Hamiltonian Eq.~(\ref{HamiltonianKAM}) can be written as 
\begin{eqnarray}
H
=
\sum_{ij\sigma}
\gamma_{ij}
{\bf P}_{i}^{\dagger}
{\bf W}
{\bf P}_{j}
c_{i\sigma}^{\dagger}c_{j\sigma}
+
\frac{J_K}{2}\sum_{i}\hat{x}_{i}\sum_{\sigma\sigma '}
S_{i}^{\sigma\sigma '}c_{i\sigma '}^{\dagger}c_{i\sigma}~, 
\label{HamiltonianKAMDMFT}
\end{eqnarray}
where we introduced the transfer matrix
\begin{eqnarray}
{\bf W}
=
\left( 
\begin{array}{lr}
t_{A} & t_{AB} \\
t_{AB} & t_{B}
\end{array}
\right)~, 
\end{eqnarray}
and the projection operators 
\begin{equation}
{\bf P}_{i}
=
\left( 
\begin{array}{c}
 \hat{x}_i \\
 \hat{y}_i  
\end{array}
\right)~, 
\end{equation}
with their conjugates
\begin{eqnarray}
{\bf P}_{i}^{\dagger}
=
\left(
\begin{array}{ccc}
\hat{x}_i &,&
\hat{y}_i
\end{array}
\right)~, 
\end{eqnarray}
where $\hat{x}_i\equiv1-\hat{y}_i$ is unity if $i$ is an $A$ site and zero otherwise. 
Here, we have implicitly mapped the initial KAM~(\ref{HamiltonianKAM}), 
characterised by a
single disordered conduction band, into a two-band effective model. 
Thereby 
each site of the underlying periodic lattice acts like being occupied 
simultaneously by 
atoms of $A$ and $B$ type. As before, the initial physical Hilbert space corresponding 
to a single kind of atom per site is recovered by introducing projection 
operators. 
They guarantee that 
a site acts either as an $A$ or $B$ atom. 
Here, we follow the DMFT formalism~[\onlinecite{reviewDMFT1}-\onlinecite{reviewDMFT2}], 
which is exact in the 
limit of a large coordination number $z$. 
Considering that the energy of the system is an
extensive quantity, this limit requires a rescaling of the hopping energies 
$t_{ij}=\gamma_{ij}{\bf P}_{i}^{\dagger}{\bf W}
{\bf P}_{j}=\tilde{t}_{ij}/\sqrt{z}$, where $\tilde{t}_{ij}$ remains finite 
(i.e., independent of $z$) when $z\to\infty$. 
From the lattice Hamiltonian 
Eq.~(\ref{HamiltonianKAMDMFT}) we obtain a local effective action for site $0$
\begin{eqnarray}
{\cal S}(\hat{x}_{0})
&=& 
-
\sum_{\sigma}
\int_{0}^{\beta}d\tau \int_{0}^{\beta}d\tau '
c_{0\sigma}^{\dagger}(\tau )
{\bf P}_{0}^{\dagger}
{\bf K}(\tau-\tau ')
{\bf P}_{0}
c_{0\sigma}(\tau ')
-
\hat{x}_{0}
\frac{J_{K}}{2}
\sum_{\sigma\sigma '}
\int_{0}^{\beta}d\tau 
S^{\sigma\sigma '}(\tau)
c_{0\sigma}^{\dagger}(\tau)
c_{0\sigma '}(\tau)~. 
\label{ActionDMFT} 
\end{eqnarray}
Here, the kernel  ${\bf K}$ is a $2\times 2$ matrix, which is a dynamical 
generalization of the Weiss field usualy introduced for a static mean-field 
approximation. The projection
operators ${\bf P}_{0}$ and ${\bf P}_{0}^{\dagger}$ select the diagonal matrix element
${\cal K}_{A}$ (respectively ${\cal K}_{B}$) of ${\bf K}$ 
depending on wether site $0$ is occupied by an $A$ or $B$ atom. 
The resulting local effective action ${\cal S}(\hat{x}_{0})$ remains 
a scalar quantity, which can
have two values, i.e., ${\cal S}(\hat{x}_{0}=1)={\cal S}_A$ and 
${\cal S}(\hat{x}_{0}=0)={\cal S}_B$. 
This is a key quantity in the DMFT procedure. 
It provides a relevant simplification since 
the local electronic and magnetic Green's functions characterizing the 
lattice Hamiltonian~(\ref{HamiltonianKAMDMFT}) can now be computed 
from ${\cal S}(\hat{x}_{0})$, which invokes local degrees of freedom only. 
Next, we determine the self-consistent relations allowing to express the
kernel 
${\bf K}$ as a function of the local Green's functions. 
Following the standard DMFT formalism, we find 
\begin{eqnarray}
{\bf K}(i\omega )
=
(i\omega+\mu){\bf I}
-\sum_{ij}
\gamma_{0i}\gamma_{j0}
{\bf W}
\left[
{\bf P}_{i}
{\bf P}_{j}^{\dagger}
G_{ij}^{(0)}(i\omega )
\right]
{\bf W}~.
\label{KernelDMFT} 
\end{eqnarray}
Here $G_{ij}^{(0)}$ is the cavity Green's function, corresponding to the 
lattice Hamiltonian~(\ref{HamiltonianKAMDMFT}), but with site $0$ excluded. 
In order to establish a self-consistent relation for the kernel, 
we perform an infinite order perturbation expansion of Green's functions
in terms of the hopping elements 
$\gamma_{ij}
{\bf P}_{i}^{\dagger}{\bf W}{\bf P}_{j}$. 
Following the DMFT scheme~[\onlinecite{reviewDMFT1}-\onlinecite{reviewDMFT2}], 
the Green's function $G_{ij}$ 
for a given distribution of sites $A$ and
$B$ is expressed as a sum of all possible paths 
$i\rightarrow i_{1}\rightarrow i_{2}\cdots i_{p}\rightarrow j$ connecting 
site $i$ to site $j$ through the sequence of structure factors $\gamma_{i_{1}i_{2}}$. 
In the limit of large $z$ we may exclude returning paths since their 
contribution is of order $1/z^{n+1}$ when $n$ is the number of returns. 
Thus, each path is factorised in terms of local dressed 
irreducible {\it scalar} propagators $\Pi_{ii}$ which contain information
about the local interactions:  
\begin{eqnarray}
G_{ij}
=
\sum_{paths}
\Pi_{ii}
\gamma_{ii_{1}}
{\bf P}_{i}^{\dagger}{\bf W}{\bf P}_{i_1}
\Pi_{i_1i_1}
\gamma_{i_{1}i_{2}}
{\bf P}_{i_{1}}^{\dagger}{\bf W}{\bf P}_{i_2}
\Pi_{i_2i_2}
\cdots
\Pi_{i_pi_p}
\gamma_{i_{p}j}
{\bf P}_{i_p}^{\dagger}{\bf W}{\bf P}_{j}
\Pi_{jj}~. 
\label{ExpansionGreenDMFT}
\end{eqnarray}
For the sake of simplicity we have dropped the explicit time dependencies. 
We will show below that, after averaging over the randomness, 
these local propagators can be related to a local self-energy. 
The large $z$ expansion Eq.~(\ref{ExpansionGreenDMFT}) is a {\it scalar} 
relation, similar to the one obtained in the usual DMFT approach. 
The only difference arises from the scalar hopping elements 
$t_{ij}=
{\bf P}_{i}^{\dagger}{\bf W}{\bf P}_{j}\gamma_{ij}$, which here are random. 
In the following we cast this relation into a $2\times 2$ matrix form, 
with a periodic effective hopping matrix ${\bf W}\gamma_{ij}$ 
between nearest neighbors.  
We define
\begin{eqnarray}
{\bf G}_{ij}
\equiv
{\bf P}_{i}G_{ij}{\bf P}_{j}^{\dagger}
=
\left(
\begin{array}{cc}
\hat{x}_i\hat{x}_j G_{ij} &
\hat{x}_i\hat{y}_j G_{ij} \\
\hat{y}_i\hat{x}_j G_{ij} &
\hat{y}_i\hat{y}_j G_{ij}
\end{array}
\right)~, 
\end{eqnarray}
and 
\begin{eqnarray}
{\bf \Pi}_{ii}
\equiv
{\bf P}_{i}\Pi_{ii}{\bf P}_{i}^{\dagger}
=
\left(
\begin{array}{cc}
\hat{x}_i \Pi_{ii} &
0 \\
0 &
\hat{y}_i \Pi_{ii}
\end{array}
\right)~. 
\end{eqnarray}
The large$-z$ expansion for the Green's function matrix ${\bf G}_{ij}$ 
is obtained by multiplying Eq.~(\ref{ExpansionGreenDMFT}) with 
the projection operators ${\bf P}_{i}^{\dagger}$ (from the left) and 
${\bf P}_{j}$ (from the right). 
After averaging with respect to
the different configurations of $A$ and $B$ sites, we find
\begin{eqnarray}
\left[
{\bf G}_{ij}
\right]
=
\sum_{paths}
\left[
{\bf \Pi}_{ii}
\gamma_{ii_{1}}
{\bf W}~
{\bf \Pi}_{i_1i_1}
\gamma_{i_{1}i_{2}}
{\bf W}~
{\bf \Pi}_{i_2i_2}
\cdots
{\bf \Pi}_{i_pi_p}
\gamma_{i_{p}j}
{\bf W}~
{\bf \Pi}_{jj}
\right]~. 
\label{ExpansionDMFTaveraged}
\end{eqnarray}
In the large$-z$ limit, we consider only direct paths connecting sites
$i$ and $j$. 
We assume now that the occupation of a given site by an $A$ or $B$ atom 
is purely random and does not depend on the configurations of the neighboring
sites. 
Therefore, in Eq.~(\ref{ExpansionDMFTaveraged}), 
each irreducible propagator matrix 
${\bf \Pi}_{ii}$ can thus be averaged separately, and we find
\begin{eqnarray}
\left[ 
{\bf G}_{ij}
\right]
=
\sum_{paths}
{\bf \Pi}_{0}
\gamma_{ii_{1}}
{\bf W}~
{\bf \Pi}_{0}
\gamma_{i_{1}i_{2}}
{\bf W}~
{\bf \Pi}_{0}
\cdots
{\bf \Pi}_{0}
\gamma_{i_{p}j}
{\bf W}~
{\bf \Pi}_{0}~, 
\label{ExpansionDMFTaveragedbis}
\end{eqnarray}
where ${\bf \Pi}_{0}\equiv 
\left[ {\bf \Pi}_{ii}\right]$. 
The matrix relation Eq.~(\ref{ExpansionDMFTaveragedbis}) 
between the average Green's 
functions, the averaged local dressed propagator, and the hopping elements 
is formally identical to a scalar expansion obtained for a regular 
periodic system within the standard DMFT formalism~[\onlinecite{reviewDMFT1}-\onlinecite{reviewDMFT2}]. 
We introduce the averaged local Green's function
\begin{eqnarray}
{\bf G}_{loc}
\equiv
\left[ {\bf G}_{ii}\right]
=
\left(
\begin{array}{cc}
x G_{A} &
0 \\
0 &
(1-x) G_{B}
\end{array}
\right)~, 
\label{ExpressionGreenlocDMFT}
\end{eqnarray}
where $G_A$ and $G_B$ are the local Green's functions corresponding to an 
$A$ site (respectively $B$ site), once averages over all the other sites configurations have been
taken. 
Using Eq.~(\ref{ExpansionDMFTaveragedbis}), 
the relation between the cavity and full Green's
function reads
\begin{eqnarray}
\left[ {\bf P}_i {\bf P}_{j}^{\dagger}G_{ij}^{(0)}\right]
=
\left[ {\bf G}_{ij}\right]
-
\left[ {\bf G}_{i0}\right]
{\bf G}_{loc}^{-1}
\left[ {\bf G}_{0j}\right]
~.  
\label{RelationcavityDMFT}
\end{eqnarray}
Averaging over a random distribution of sites $A$ and $B$ restores the
translation symmetry of the underlying lattice. The averaged Green's  
function matrices are thus periodic in space, and we can define their Fourier 
transforms
\begin{eqnarray}
{\bf G}_{{\bf k}}
\equiv 
\sum_{ij}e^{-i{\bf k}.({\bf R}_{i}-{\bf R}_{j})}
\left[ {\bf G}_{ij} \right]~. 
\end{eqnarray}
From Eq.~(\ref{ExpansionDMFTaveragedbis}), we find that 
the Green's functions are characterised by a  local 
$2\times 2$ self-energy matrix ${\bf \Sigma}$  
\begin{eqnarray}
{\bf G}_{{\bf k}}^{-1}(i\omega)
=
(i\omega +\mu){\bf I}
-
{\bf \Sigma}(i\omega)
-
{\bf W}\gamma_{{\bf k}}~, 
\label{ExpressionGreenkDMFT} 
\end{eqnarray}
where ${\bf \Sigma}$ is related to the averaged local propagator by the 
matrix identity 
${\bf \Pi}_{0}^{-1}(i\omega)=
(i\omega +\mu){\bf I}-{\bf \Sigma}(i\omega)$. 
The matrix elements of ${\bf \Sigma}$ can be expressed in terms 
of the local average Green's function matrix 
${\bf G}_{loc}$ by taking the inverse of 
\begin{eqnarray}
{\bf G}_{loc}(i\omega)
=
\sum_{{\bf k}}
{\bf G}_{{\bf k}}(i\omega)~. 
\label{ConsistencyGreenlocDMFT}
\end{eqnarray}
Finally, using the relation~(\ref{RelationcavityDMFT}) 
for the cavity Green's function, 
together with the expression~(\ref{KernelDMFT}) for the kernel, we find
\begin{eqnarray}
{\bf K}(i\omega )
=
{\bf \Sigma}(i\omega)
+
{\bf G}_{loc}^{-1}(i\omega)~. 
\label{ConsistencyKernelDMFT}
\end{eqnarray}
Equations~(\ref{ExpressionGreenkDMFT}, 
\ref{ConsistencyGreenlocDMFT}, 
\ref{ConsistencyKernelDMFT}) 
provide a self-consistent relation between the matrix kernel ${\bf K}$ 
and the averaged local Green's function matrix ${\bf G}_{loc}$: 
\begin{eqnarray}
{\bf G}_{loc}(i\omega)
=
\sum_{{\bf k}}
\left( 
(i\omega +\mu){\bf I}
-
{\bf K}(i\omega )
+
{\bf G}_{loc}^{-1}(i\omega)
-
{\bf W}\gamma_{{\bf k}}
\right)^{-1}~. 
\label{SelfconsistencyKGloc}
\end{eqnarray}
In turn, the local Green's functions $G_A$ and $G_B$ invoked in the 
definition~(\ref{ExpressionGreenlocDMFT}) of ${\bf G}_{loc}$ can be computed 
for a given kernel ${\bf K}$, by considering the cases $\hat{x}_0=0$ and 
$\hat{x}_0=1$ in the local effective action Eq.~(\ref{ActionDMFT}). 
Since the effective action ${\cal S}_{B}$ 
on sites $B$ is quadratic in terms of electronic opperators, 
we obtain an explicit expression for $G_B$: 
\begin{eqnarray}
G_B(i\omega)={\cal K}_{B}^{-1}(i\omega)~. 
\end{eqnarray}
The local Green's functions $G_A$ is obtained from the local 
effective action on an $A$ atom: 
\begin{eqnarray}
{\cal S}_{A}
&=& 
-
\sum_{\sigma}
\int_{0}^{\beta}d\tau \int_{0}^{\beta}d\tau '
c_{0\sigma}^{\dagger}(\tau )
{\cal K}_{A}(\tau-\tau ')
c_{0\sigma}(\tau ')
-
\frac{J_K}{2}
\sum_{\sigma\sigma '}
\int_{0}^{\beta}d\tau 
S^{\sigma\sigma '}(\tau)
c_{0\sigma}^{\dagger}(\tau)
c_{0\sigma '}(\tau)~. 
\label{ActionDMFTsiteA} 
\end{eqnarray}
Appart from the self-consistent relation~(\ref{SelfconsistencyKGloc}), 
which can be treated using analytic (and eventually numerical)
simple calculations, the main difficulty consists in computing $G_A$ from 
the local effective action~(\ref{ActionDMFTsiteA}). 
Even if the initial difficulty of studying a lattice model has been 
consequently reduced into a single site effective model, this issue remains 
a many body problem. The Kondo interaction part has to be considered using 
a numerical scheme or appropriate analytical approximations. 

Once a self-consistent solution is obtained for ${\bf K}$ and ${\bf G}_{loc}$, 
the ${\bf k}-$dependent correlation functions for the conduction electrons can
be obtained using 
Eqs.~(\ref{ExpressionGreenkDMFT}, 
\ref{ConsistencyKernelDMFT}). 
Here, we describe the system with two bands of conduction electrons $A$ and
$B$, whose correlations are characterised by the $2\times 2$ matrix 
${\bf G}_{{\bf k}}$. Invoking the identity 
$\hat{x}_{i}\hat{x}_{j}+\hat{x}_{i}\hat{y}_{j}+\hat{y}_{i}\hat{x}_{j}+\hat{y}_{i}\hat{y}_{j}=1$, 
the physical single band average Green's functions 
$\left[ G_{ij}\right]$ can be obtained by adding the four matrix elements
of $\left[ {\bf G}_{ij}\right]$. As a consequence, the 
${\bf k}-$dependent average correlation function for the physical single band
of conduction electrons is also obtained by adding the four matrix elements of 
${\bf G}_{{\bf k}}$.

\subsection{Equivalence of the CPA and the DMFT}
The equivalence of the dynamical CPA and the DMFT was previously
proven by Kakehashi on general grounds~[\onlinecite{Kakehashi}]. 
As discussed before by applying a $3\times 3$ matrix CPA approach we were able
to describe the important dynamical aspects of the spins of the $A$
sites. Therefore it is reassuring that we can demonstrate the equivalence
of the $3\times3$ matrix CPA approach with corresponding DMFT equations, when
we integrate over the $f-$electron degrees of freedom in the CPA approach and
make a mean-field approximation within the DMFT approach. 

\subsubsection{Expression of the CPA equations using a $2\times 2$ matrix formalism}
For a demonstration of the equivalence of the two methods we start from a
modified version of  Eqs.~(\ref{DefinitionSigmaCPA}-
\ref{RelationSigmaBCPA}). 
It is easy to show that after some algebraic modifications the following
relations can be derived from these equations: 
\begin{eqnarray}
G_{A}^{ff}(i\omega)
&=&
{1\over{i\omega+\lambda}}
+
{{r^2}\over{(i\omega+\lambda)^{2}}}
G_{A}^{cc}(i\omega)~, 
\label{ExpressionGreenAffCPA}
\\
G_{A}^{fc}(i\omega)
&=&
{{-r}\over{i\omega+\lambda}}
G_{A}^{cc}(i\omega)~. 
\label{ExpressionGreenAfcCPA}
\end{eqnarray}
The self-consistent CPA 
Eqs.~(\ref{DefinitionSigmaCPA}, 
\ref{RelationGreenreciprocalCPA}, 
\ref{RelationSigmaACPA}, 
\ref{RelationSigmaBCPA}) 
can be cast into the form
\begin{eqnarray}
xG_{A}^{cc}(i\omega)
&=&
\sum_{{\bf k}}
{1\over{
\Delta_{{\bf k}}
(i\omega)}}
\left(
i\omega+\mu-\Sigma_{B}^{CPA}(i\omega)-t_{B}\gamma_{{\bf k}}
\right)~, 
\label{RelationGreenACPA}\\
(1-x)G_{B}^{cc}(i\omega)
&=&
\sum_{{\bf k}}
{
1\over{
\Delta_{{\bf k}}
(i\omega)}}
\left(
i\omega+\mu
-\Sigma_{K}(i\omega)
-
\Sigma_{A}^{CPA}(i\omega)
-t_{A}\gamma_{{\bf k}}
\right)~, 
\label{RelationGreenBCPA}
\end{eqnarray}
where 
\begin{eqnarray}
\Delta_{{\bf k}}(i\omega) 
&=&
\left(
i\omega+\mu
-\Sigma_{K}(i\omega)
-
\Sigma_{A}^{CPA}(i\omega)
-t_{A}\gamma_{{\bf k}}
\right)
\left(
i\omega+\mu-\Sigma_{B}^{CPA}(i\omega)-t_{B}\gamma_{{\bf k}}
\right)
-
\left(
-\sigma_{2}(i\omega) -t_{AB}\gamma_{{\bf k}}
\right)^{2}~, 
\label{RelationDeltaCPA} \\
&&\nonumber\\
\sigma_{2}(i\omega)
&=&
-t_{AB}
{
{1-2x
+x(i\omega+\mu-\Sigma_{K}(i\omega))G_{A}(i\omega)
-(1-x)(i\omega+\mu)G_{B}(i\omega)
}
\over{
xt_{A}G_{A}(i\omega)
-(1-x)t_{B}G_{B}(i\omega)
}}~. 
\end{eqnarray}
Here we have set
\begin{eqnarray}
\Sigma_{A}^{CPA}(i\omega)
&=&
-{{1-x}\over{x}}
{1\over{G_{A}^{cc}(i\omega)}}~, 
\label{ExpressionSigmaACPA}
\\
\Sigma_{B}^{CPA}(i\omega)
&=&
-{{x}\over{1-x}}
{1\over{G_{B}^{cc}(i\omega)}}~.  
\label{ExpressionSigmaBCPA}
\end{eqnarray}
We have also introduced the definition
\begin{eqnarray}
\Sigma_{K}(i\omega)
\equiv
{{r^2}\over{i\omega+\lambda}}~. 
\label{ExpressionSigmaKCPA}
\end{eqnarray}
A complete resolution of the model is obtained by the following 
self-consistent scheme: \\
{\it (i)}~~~Calculate $G_{A}^{cc}$ and $G_{B}^{cc}$ 
from
Eqs.~(\ref{RelationGreenACPA}, 
\ref{RelationGreenBCPA}, 
\ref{RelationDeltaCPA}, 
\ref{ExpressionSigmaACPA}, 
\ref{ExpressionSigmaBCPA}, 
\ref{ExpressionSigmaKCPA}), 
as function of the mean-field parameters $r$, $\lambda$ and $\mu$. \\
{\it (ii)}~~Calculate $G_{A}^{fc}$ and $G_{A}^{ff}$ by using 
Eqs.~(\ref{ExpressionGreenAffCPA}, 
\ref{ExpressionGreenAfcCPA}). \\
{\it (iii)}~Optimise the parameters $r$, $\lambda$ and $\mu$ so as 
to satisfy the mean-field  
Eqs.~(\ref{EquationlargeN1}, 
\ref{EquationlargeN2}, 
\ref{EquationlargeN3}).

\subsubsection{DMFT and mean-field approximation for the Kondo term}
A complete resolution of the matrix DMFT self-consistent relations 
requires an impurity solver, in order to compute the local electronic 
Green's functions $G_A$ related to the local effective 
action ${\cal S}_{A}$ given by Eq.~(\ref{ActionDMFTsiteA}). 
In order to demonstrate the formal equivalence between the matrix-DMFT 
and the matrix-CPA approaches, we use the mean-field 
approximation for the impurity solver. Before, we define the local self-energy 
due to the Kondo interaction on sites $A$: 
\begin{eqnarray}
\Sigma_{K}(i\omega)
\equiv
{\cal K}_{A}(i\omega)
-G_{A}^{-1}(i\omega)
\end{eqnarray}
Since there is no local interaction on sites $B$, we have 
${G}_{B}(i\omega)={\cal K}_{B}^{-1}(i\omega)$. 
The relation~(\ref{ConsistencyKernelDMFT}) can be expressed as 
\begin{eqnarray}
{\bf \Sigma}(i\omega)
=
\left(
\begin{array}{ll}
\Sigma_{K}(i\omega)
+
\Sigma_{A}^{CPA}(i\omega)
&
{\cal K}_{AB}(i\omega)
\\
{\cal K}_{AB}(i\omega)
&
\Sigma_{B}^{CPA}(i\omega)
\end{array}
\right)
\label{ExpressionSigmaDMFT}
\end{eqnarray}
with
\begin{eqnarray}
\Sigma_{A}^{CPA}(i\omega)
&=&
((1-x)/x)G_{A}^{-1}(i\omega)
\label{ExpressionSigmaADMFT}
\\
\Sigma_{B}^{CPA}(i\omega)
&=&
(x/(1-x))G_{B}^{-1}(i\omega)
\label{ExpressionSigmaBDMFT}
\end{eqnarray}
These expressions are identical to 
Eqs.~(\ref{ExpressionSigmaACPA}, 
\ref{ExpressionSigmaBCPA}) obtained within the 
matrix-CPA approach. 
Here, as within the CPA approach, the off-diagonal self-energy 
${\cal K}_{AB}$ (denoted before $\sigma_{2}$) is determined 
by requiring the vanishing of the off-diagonal elements of 
${\bf G}_{loc}$ in Eq.~(\ref{ExpressionGreenlocDMFT}). 
In analogy to Eq.~(\ref{Expressionsigma2CPA}), we find
\begin{eqnarray}
{\cal K}_{AB}(i\omega)
=\sigma_{2}(i\omega)
=
-t_{AB}
{
{1-2x
+x(i\omega+\mu-\Sigma_{K}(i\omega))G_{A}(i\omega)
-(1-x)(i\omega+\mu)G_{B}(i\omega)
}
\over{
xt_{A}G_{A}(i\omega)
-(1-x)t_{B}G_{B}(i\omega)
}}~. 
\label{Expressionsigma2DMFT}
\end{eqnarray}
Combining 
Eq.~(\ref{ExpressionSigmaDMFT}) with the self-consistent relations 
Eqs.~(\ref{ExpressionGreenlocDMFT}, 
\ref{ExpressionGreenkDMFT}, 
\ref{ConsistencyGreenlocDMFT}) we find 
\begin{eqnarray}
xG_{A}(i\omega)
&=&
\sum_{{\bf k}}
{1\over{
\Delta_{{\bf k}}
(i\omega)}}
\left(
i\omega+\mu-\Sigma_{B}^{CPA}(i\omega)-t_{B}\gamma_{{\bf k}}
\right)~, 
\label{RelationGreenADMFT}
\\
(1-x)G_{B}(i\omega)
&=&
\sum_{{\bf k}}
{
1\over{
\Delta_{{\bf k}}
(i\omega)}}
\left(
i\omega+\mu
-\Sigma_{K}(i\omega)
-
\Sigma_{A}^{CPA}(i\omega)
-t_{A}\gamma_{{\bf k}}
\right)~, 
\label{RelationGreenBDMFT}
\end{eqnarray}
with
\begin{eqnarray}
\Delta_{{\bf k}}(i\omega) 
&=&
\left(
i\omega-\mu
-\Sigma_{K}(i\omega)
-
\Sigma_{A}^{CPA}(i\omega)
-t_{A}\gamma_{{\bf k}}
\right)
\left(
i\omega-\mu-\Sigma_{B}^{CPA}(i\omega)-t_{B}\gamma_{{\bf k}}
\right)
-
\left(
-\sigma_{2}(i\omega) -t_{AB}\gamma_{{\bf k}}
\right)^{2}~, 
\label{RelationDeltaDMFT}
\end{eqnarray}
which are formally equivalent to the relations 
Eqs~(\ref{RelationGreenACPA}, 
\ref{RelationGreenBCPA}, 
\ref{RelationDeltaCPA}) obtained from the matrix form of the CPA approach. 

The matrix-DMFT approach developed here is performed without
any approximation concerning the local Kondo interaction. 
An impurity solver is required in order to calculate the local Green's 
functions from the local effective action ${\cal S}_{A}$ defined in 
Eq.~(\ref{ActionDMFTsiteA}), and then to compute the Kondo self-energy
$\Sigma_{K}$. 
For example, the mean-field approximation can be performed 
as described in the previous section (CPA),
leading to the same set of saddle point relations as 
Eqs.~(\ref{EquationlargeN1}, 
\ref{EquationlargeN2}, 
\ref{EquationlargeN3}). 
This method, developped in the framework of a Kondo Alloy Model 
can be generalised to other alloy models with strong 
local correlations.

\subsection{Bethe lattice with infinite coordination}
The DMFT formalism described in the previous section is exact in the
limit of an infinite coordination number $z$. It can be applied to any 
underlying periodic lattice, which has to be defined by 
its structure factor $\gamma_{{\bf k}}$. 
In order to study numerically the Kondo Alloy Model defined by 
the Hamiltonian Eq.~(\ref{HamiltonianKAM}), it appears as very convenient 
to consider a Bethe lattice. 
For a similar approach applied to ferromagnetic semiconductors see 
Ref~[\onlinecite{DasSarma}]. 
In this specific case, the self-consistent equations are much simpler, 
and the general physical properties of the system are preserved. 

Applying the DMFT formalism described in the previous 
section to a Bethe lattice, we obtain a local effective action for 
the two kind of sites 
\begin{eqnarray}
{\cal S}_{A}
&=& 
-\sum_{\sigma}\int_{0}^{\beta}d\tau \int_{0}^{\beta}d\tau '
c_{A\sigma}^{\dagger}(\tau )
{\cal K}_{A}(\tau-\tau ')c_{A\sigma}(\tau ')
-
\frac{J_K}{2}
\sum_{\sigma\sigma '}
\int_{0}^{\beta}d\tau 
S^{\sigma\sigma '}(\tau)
c_{A\sigma}^{\dagger}(\tau)
c_{A\sigma '}(\tau)~,
\label{ActionABethelattice}
\\
{\cal S}_{B}
&=& 
-\sum_{\sigma}\int_{0}^{\beta}d\tau \int_{0}^{\beta}d\tau '
c_{B\sigma}^{\dagger}(\tau )
{\cal K}_{B}(\tau-\tau ')c_{B\sigma}(\tau ')~. 
\label{ActionBBethelattice}
\end{eqnarray}
They are formaly equivalent to the compact expression Eq.~(\ref{ActionDMFT}). 
The main simplification obtained by considering a Bethe lattice rests on the 
fact that the cavity Green's functions involved in Eq.~(\ref{KernelDMFT}) 
can now be replaced by local full Green's functions. 
This procedure is exact in the limit of a large coordination number $z$. 
The Bethe lattice self-consistent relations for the Kernels ${\cal K}_{A}$ and 
${\cal K}_{B}$ are thus 
\begin{eqnarray}
{\cal K}_{A}(i\omega )
&=&
i\omega+\mu
-x\tilde{t}_{A}^{2}G^{cc}_{A}(i\omega)
-(1-x)\tilde{t}_{AB}^{2}G^{cc}_{B}(i\omega) 
\label{BainABethelattice}
\\
{\cal K}_{B}(i\omega )
&=&
i\omega+\mu
-x\tilde{t}_{AB}^{2}G^{cc}_{A}(i\omega)
-(1-x)\tilde{t}_{B}^{2}G^{cc}_{B}(i\omega)~, 
\label{BainBBethelattice}
\end{eqnarray}
where, in the large $z$ limit, the nearest-neightbor hoppings have been 
rescaled: $t_{A}\equiv \tilde{t}_{A}/\sqrt{z}$, with similar definitions for 
$\tilde{t}_{B}$ and $\tilde{t}_{AB}$. 
We then apply the mean-field approximation, 
described in the first section, as an impurity solver for the 
local effective action ${\cal S}_{A}$ of an $A$ site. 
Within mean-field approximation, the effective actions 
Eqs.~(\ref{ActionABethelattice}, 
\ref{ActionBBethelattice}) are quadratic and the local Green's 
functions can in turn be expressed explicitly as functions of the 
kernels ${\cal K}_{A}$ and ${\cal K}_{B}$ 
\begin{eqnarray}
\left (
\begin{array}{cc}
G^{ff}_{A}(i\omega ) & G^{fc}_{A} (i\omega )\\
G^{cf}_{A}(i\omega )& G^{cc}_{A}(i\omega ) 
\end{array}
\right )
&=& 
{\left (
\begin{array}{cc}
i\omega +\lambda  & r \\
r & {\cal K}_{A}(i\omega) 
\end{array}
\right )
}^{-1}~, \\
&&\nonumber\\
G^{cc}_{B}(i\omega )
&=&
{\cal K}_{B}^{-1}(i\omega)~. 
\end{eqnarray}
Together with Eqs.~(\ref{BainABethelattice}, 
\ref{BainBBethelattice}) and with
the mean-field equations 
Eqs.~(\ref{EquationlargeN1}, 
\ref{EquationlargeN2}, 
\ref{EquationlargeN3}) we have a complete set of self-consistent 
relations for the local Green's functions and the effective 
parameters $r$, $\lambda$ and $\mu$.

\section{Applications of the formalism \label{sectionresults}}

\subsection{Off-diagonal randomness: non-magnetic random alloy 
\label{subsectionoffdiagonal}}
\subsubsection{Formalism}
In this section we consider off-diagonal randomness, i.e., hopping 
matrix elements. 
The model is defined by the Hamiltonian Eq.~(\ref{HamiltonianKAM}) 
without the Kondo interaction. 
This is a standard situation for the CPA and we discuss this case here only
because we want to combine it later with the Kondo problem. 
We know that the CPA misses certain localization effects. Their importance in
connexion with Kondo effect has been discussed in Ref.~[\onlinecite{Vojta}]. 
Since the spin components are decoupled, the system corresponds to 
a random tight-binding model of conduction electrons, 
identical to the one considered in Refs.~[\onlinecite{matrixCPABEB1}-\onlinecite{matrixCPABEB2}].  
Thus $G_{ij}(\tau-\tau')\equiv 
-\langle T_{\tau}c_{i}(\tau)c_{j}^{\dagger}(\tau')\rangle$ 
is the electron Green's function defined for imaginary time. 
Since here we do not consider the Kondo interaction, the self-consistent 
equation for the averaged Green's function 
can equivalently be obtained 
either from the matrix form of the CPA approach 
(Eqs.~(\ref{RelationGreenACPA}, 
\ref{RelationGreenBCPA}, 
\ref{RelationDeltaCPA}, 
\ref{ExpressionSigmaACPA}, 
\ref{ExpressionSigmaBCPA})) 
or from the matrix DMFT approach 
(Eqs.~\ref{ExpressionSigmaADMFT}, 
\ref{ExpressionSigmaBDMFT}, 
\ref{RelationGreenADMFT}, 
\ref{RelationGreenBDMFT}, 
\ref{RelationDeltaDMFT}). 
In both cases the Kondo self-energy $\Sigma_{K}=0$. 
We find 
\begin{eqnarray}
xG_{A}(i\omega)
&=&
\sum_{{\bf k}}
{1\over{
\Delta_{{\bf k}}
(i\omega)}}
\left(
i\omega-\Sigma_{B}^{CPA}(i\omega)-t_{B}\gamma_{{\bf k}}
\right)~, 
\label{RelationGreenATB}
\\
(1-x)G_{B}(i\omega)
&=&
\sum_{{\bf k}}
{
1\over{
\Delta_{{\bf k}}
(i\omega)}}
\left(
i\omega
-
\Sigma_{A}^{CPA}(i\omega)
-t_{A}\gamma_{{\bf k}}
\right)~, 
\label{RelationGreenBTB}
\end{eqnarray}
where 
\begin{eqnarray}
\Delta_{{\bf k}}(i\omega) 
&=&
\left(
i\omega
-
\Sigma_{A}^{CPA}(i\omega)
-t_{A}\gamma_{{\bf k}}
\right)
\left(
i\omega-\Sigma_{B}^{CPA}(i\omega)-t_{B}\gamma_{{\bf k}}
\right)
-
\left(
-\sigma_{2}(i\omega) -t_{AB}\gamma_{{\bf k}}
\right)^{2}~, 
\label{RelationDeltaTB}
\end{eqnarray}
with
\begin{eqnarray}
\Sigma_{A}^{CPA}(i\omega)
&=&
((1-x)/x)G_{A}^{-1}(i\omega)~, 
\label{ExpressionSigmaATB}
\\
\Sigma_{B}^{CPA}(i\omega)
&=&
(x/(1-x))G_{B}^{-1}(i\omega)~, 
\label{ExpressionSigmaBTB}
\end{eqnarray}
and
\begin{eqnarray}
\sigma_{2}(i\omega)
=
-t_{AB}
{
{1-2x
+ i\omega (x G_{A}(i\omega)
-(1-x)G_{B}(i\omega))
}
\over{
xt_{A}G_{A}(i\omega)
-(1-x)t_{B}G_{B}(i\omega)
}}~. 
\label{Expressionsigma2TB}
\end{eqnarray}
Here $G_{A}$ and $G_{B}$ are the local Green's functions $G_{ii}$ 
on a site $i$ of kind $A$ (or $B$ respectively), obtained by averaging 
over all the
other site configurations $A$ or $B$. 
For the sake of simplicity, we drop the chemical potential $\mu$. 
This convention implies that the Fermi level energy is zero 
when the electron band is half-filled. 

We define the local density of electronic states (DOS) associated with 
$G_{A}$ and $G_{B}$
\begin{eqnarray}
\rho_{A/B}(\omega)\equiv -(1/\pi){\cal I}m G_{A/B}(\omega+i0^{+})~, 
\end{eqnarray}
and the averaged local DOS 
\begin{eqnarray}
\rho(\omega)=x\rho_{A}(\omega)+(1-x)\rho_{B}(\omega)~. 
\end{eqnarray}
Here, within CPA or DMFT, all the local DOS's of $A$ sites are the same while a
more accurate treatment would also show a spread there. 
It can be obtained by randomizing the distribution of the hopping matrix
elements $t_{A}$, $t_{B}$ and $t_{AB}$. This could be done within the present
DMFT formalism. 
By construction, $\rho_{A}$, $\rho_{B}$, and $\rho$ have each a total spectral
weight of unity, and $A$ ($B$) atoms contribute with a weight $x$ 
($1-x$) 
to the averaged DOS $\rho$. 

We didn't find a general analytic solution for this set of equations, 
so that a numerical evaluation is required. 
Nevertheless, a dimensionless ratio emerges from these expressions
\begin{eqnarray}
\alpha 
= t_{AB}/\sqrt{t_{A}t_{B}}~, 
\label{Definitionalpha}
\end{eqnarray}
which compares the energy characterising the hopping of electrons between 
sites $A$ and $B$ with the hopping energies within an $A$ and a $B$ sublattice. 
Intuitively, if $\alpha>1$, the electronic levels of lowest energy will be 
dominated by hopping between $A-B$ neighboring sites. 
In the opposite case of $\alpha <1$, hopping within pure $A$ or 
pure $B$ sublattices dominates. 

\subsubsection{Numerical results: local density of states (DOS)}
Choosing a Bethe tree structure as underlying lattice (this corresponds to a 
semi-elliptic DOS for the pure $A$ or pure $B$ system), 
we have computed $\rho_{A}(\omega)$ and $\rho_{B}(\omega)$ numerically for 
different values of the hopping elements. 
For the purpose of simplification, we present here some 
numerical results obtained for $t_{A}=t_{B}\equiv t$ only. In the following, all
energies are expressed in units of $\tilde{t}\equiv t\sqrt{z}$, where $z$
is the coordination number of the lattice, and different values of 
$\alpha=t_{AB}/t$ are considered. 

\begin{figure}[!ht]
\includegraphics[angle=0,width=0.32\columnwidth]{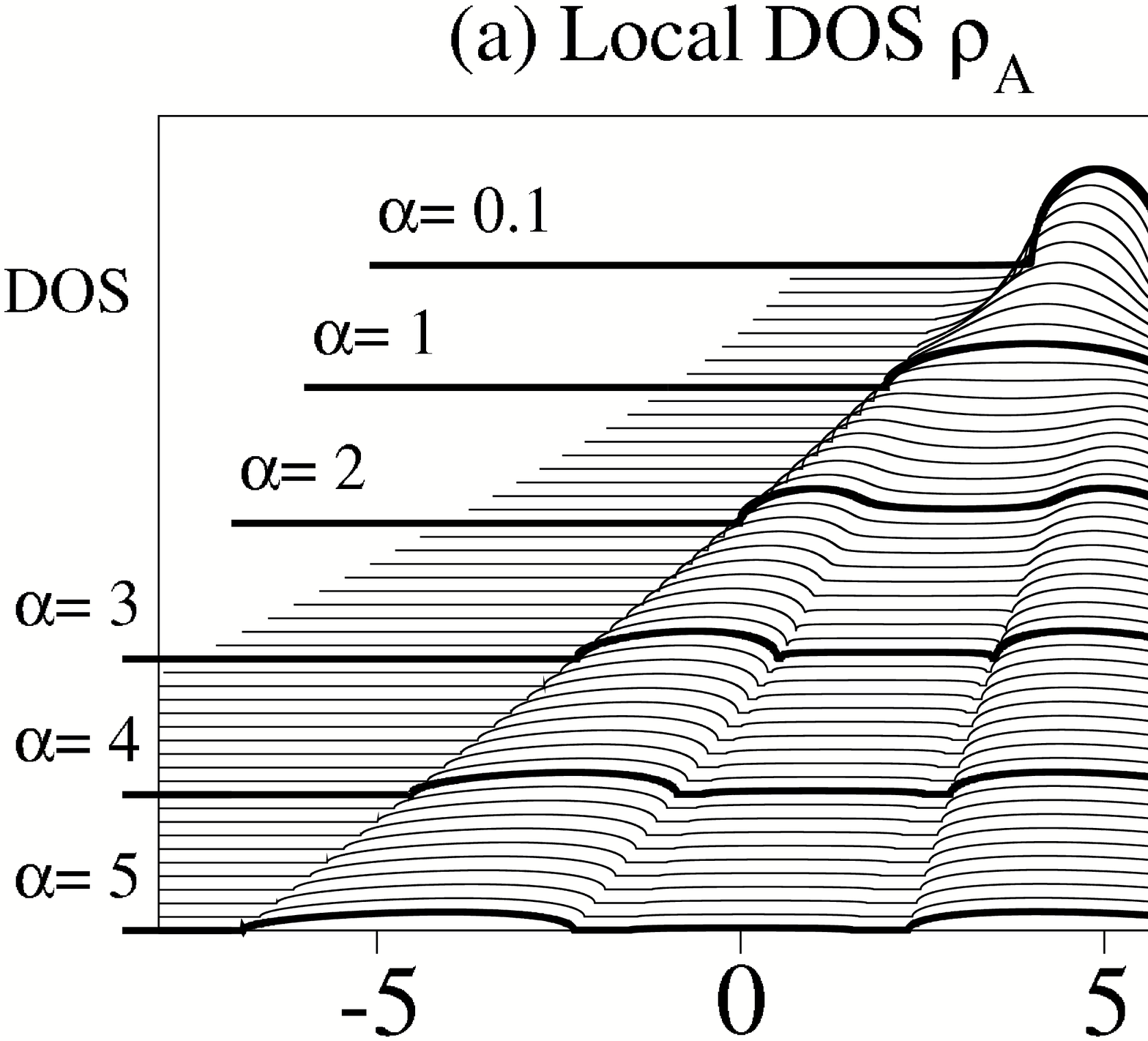}
\includegraphics[angle=0,width=0.32\columnwidth]{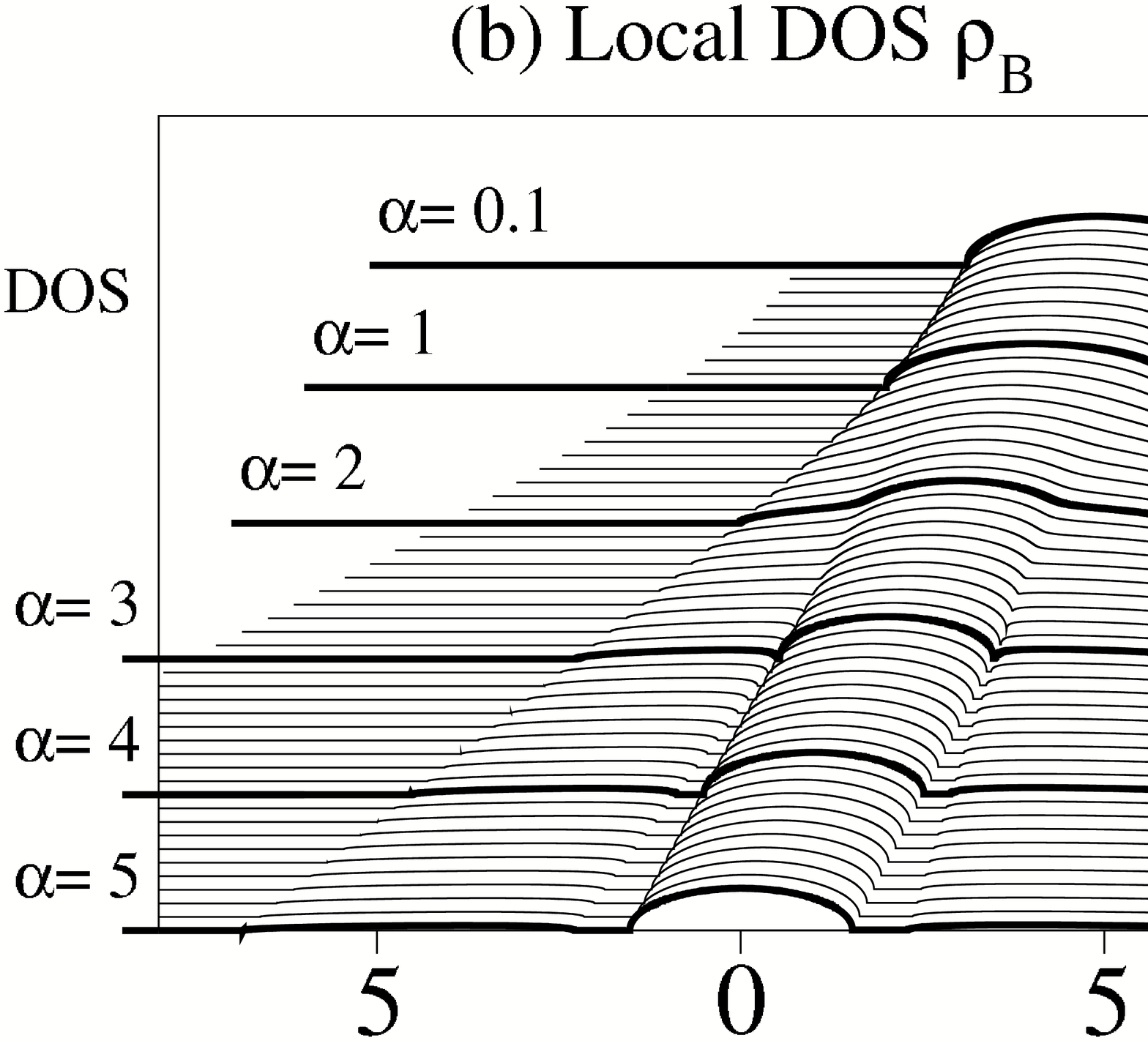}
\includegraphics[angle=0,width=0.32\columnwidth]{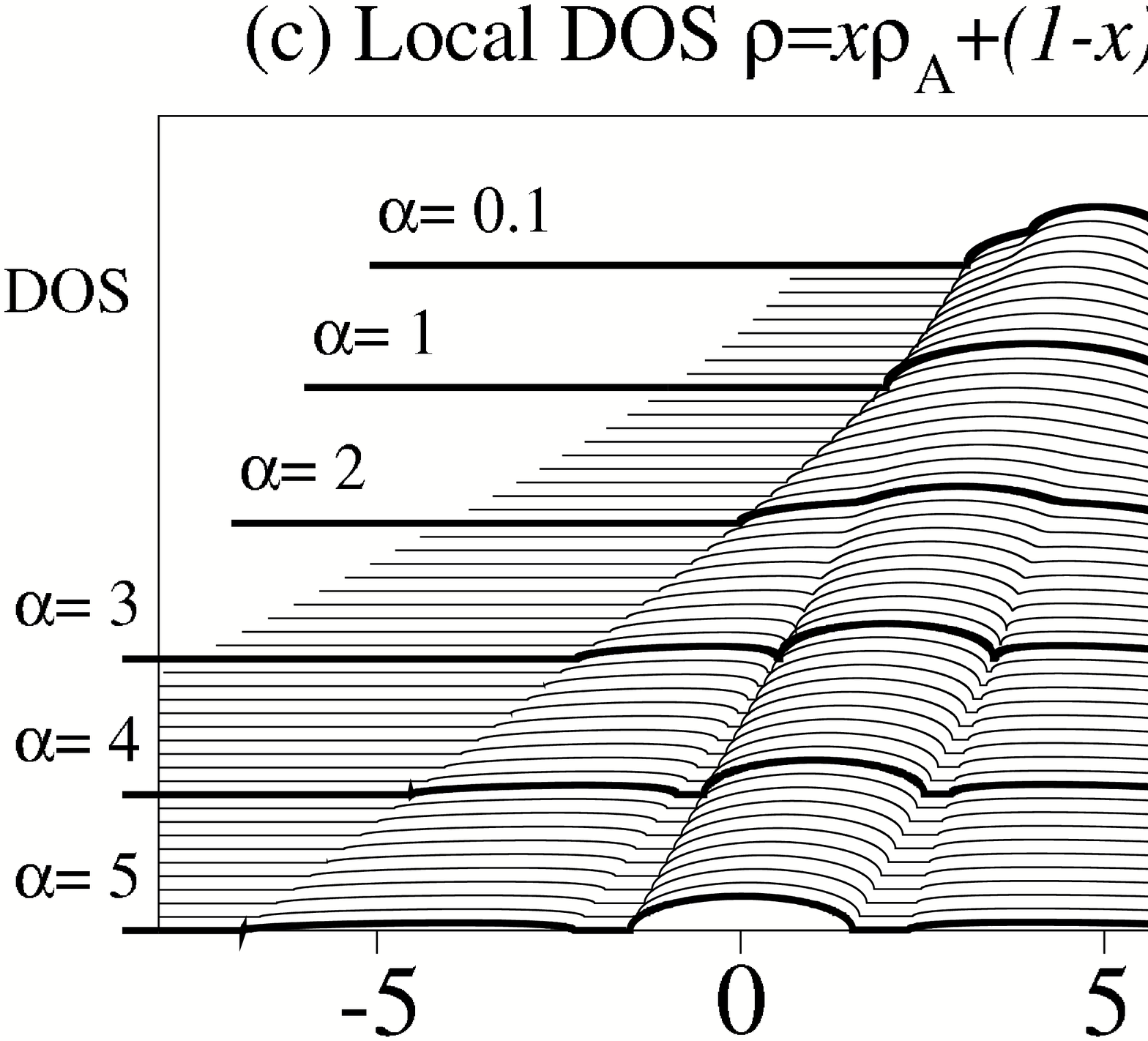}
\caption{
(a) 
Local DOS $\rho_{A}(\omega)$ on a site $A$. 
(b) 
Local DOS $\rho_{B}(\omega)$ on a site $B$. 
(c) 
Average local DOS 
$\rho(\omega)=x\rho_{A}(\omega)+(1-x)\rho_{B}(\omega)$. 
The curves correspond to a concentration $x=20\%$. 
Each plot corresponds to a fixed value of $\alpha=t_{AB}/t$, which
varies from $0.1$ (top) to $5$ (down) in steps of $0.1$. The curves have been 
shifted both vertically and horizontally, with numerical 
scales written explicitly only 
for $\alpha=0.1$ (vertical direction), and $\alpha=5$ (horizontal axis). 
The energy is in units of $\tilde{t}=t_{A}\sqrt{z}=t_{B}\sqrt{z}$. 
}
\label{FigurelocaldosrandomTB}
\end{figure}

Figure~\ref{FigurelocaldosrandomTB} depicts the effect of off-diagonal 
randomness on the local DOS's $\rho_{A}$, $\rho_{B}$, and $\rho$. 
The plots presented here are obtained for a concentration $x=20\%$ of 
atoms of kind $A$. Qualitatively similar behavior is found 
for different concentrations. 
For $\alpha=1$ ($t_{AB}=t_{A}=t_{B}$), 
our numerical results recover the semi-elliptic 
DOS corresponding to a regular tight-binding model. 
In the regime $\alpha <1$, the DOS remains semi-elliptic like, with 
a rescaled bandwidth (see also FIG.~\ref{FigureSchemasdostAB}~(b)). 
In the regime $\alpha>1$, the DOS $\rho_A$ corresponding to the minority
atoms splits into two satellites peaks, centered around the energy 
$\pm t_{AB}$, while the DOS $\rho_{B}$ corresponding to the majority 
atoms shows both a two satellites peak structure and a 
coherent peak around $\omega=0$. The latter is reminiscent of the semi-elliptic 
one obtained in the absence of disorder ($\alpha=1$). 
As a consequence, the averaged local DOS $\rho$ is also characterized by a
central peak and two satellites. 

\begin{figure}[!ht]
\includegraphics[angle=0,width=0.5\columnwidth]{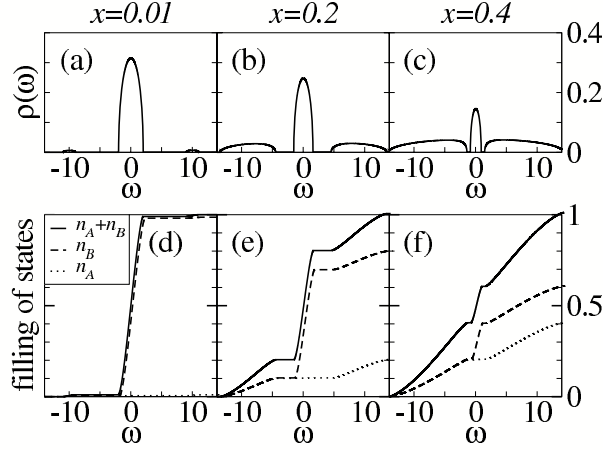}
\caption{(a-b-c) Average local DOS 
$\rho(\omega)=x\rho_{A}(\omega)+(1-x)\rho_{B}(\omega)$ (top). 
(d-e-f) Filling of states on $A$ sites (dotted line), $B$ sites (dashed line), 
and 
both $A$ and $B$ sites (solid line), as a function of the Fermi level energy. 
Numerical results for $\alpha=10$. The energy is in units of 
$\tilde{t}=t_{A}\sqrt{z}=t_{B}\sqrt{z}$. 
The concentration $x$ of $A$ atoms is equal to $0.01$ ((a)and (d)), 
$0.2$ ((b) and (e)), and $0.4$ ((c) and (f)). 
The curves corresponding to $x=0.6;0.8;0.99$ can be deduced by inverting
$A$ and $B$. }
\label{dosfreetab10.ps}
\end{figure}

\begin{figure}[!ht]
\includegraphics[angle=270,width=0.4\columnwidth]{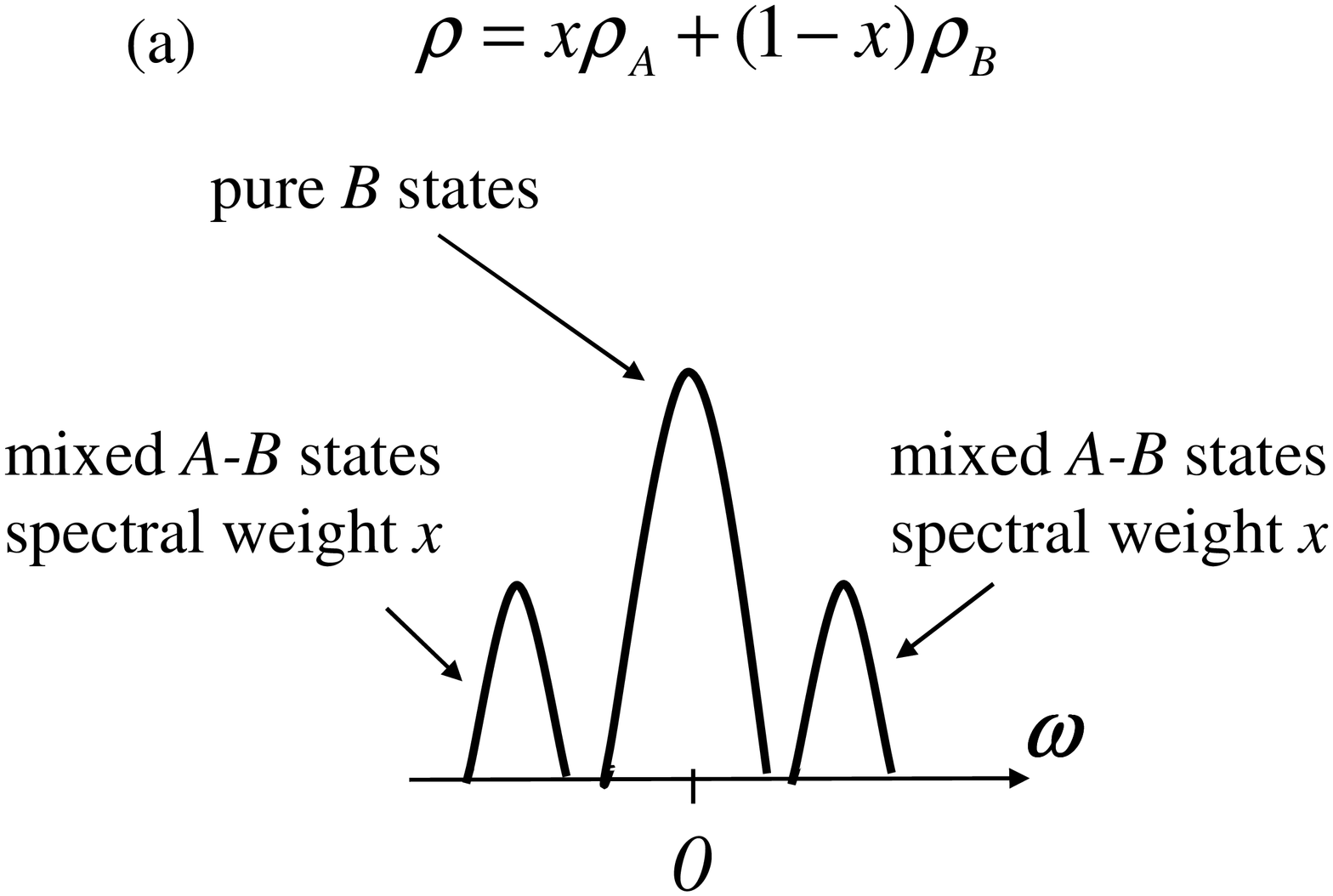}
\includegraphics[angle=270,width=0.4\columnwidth]{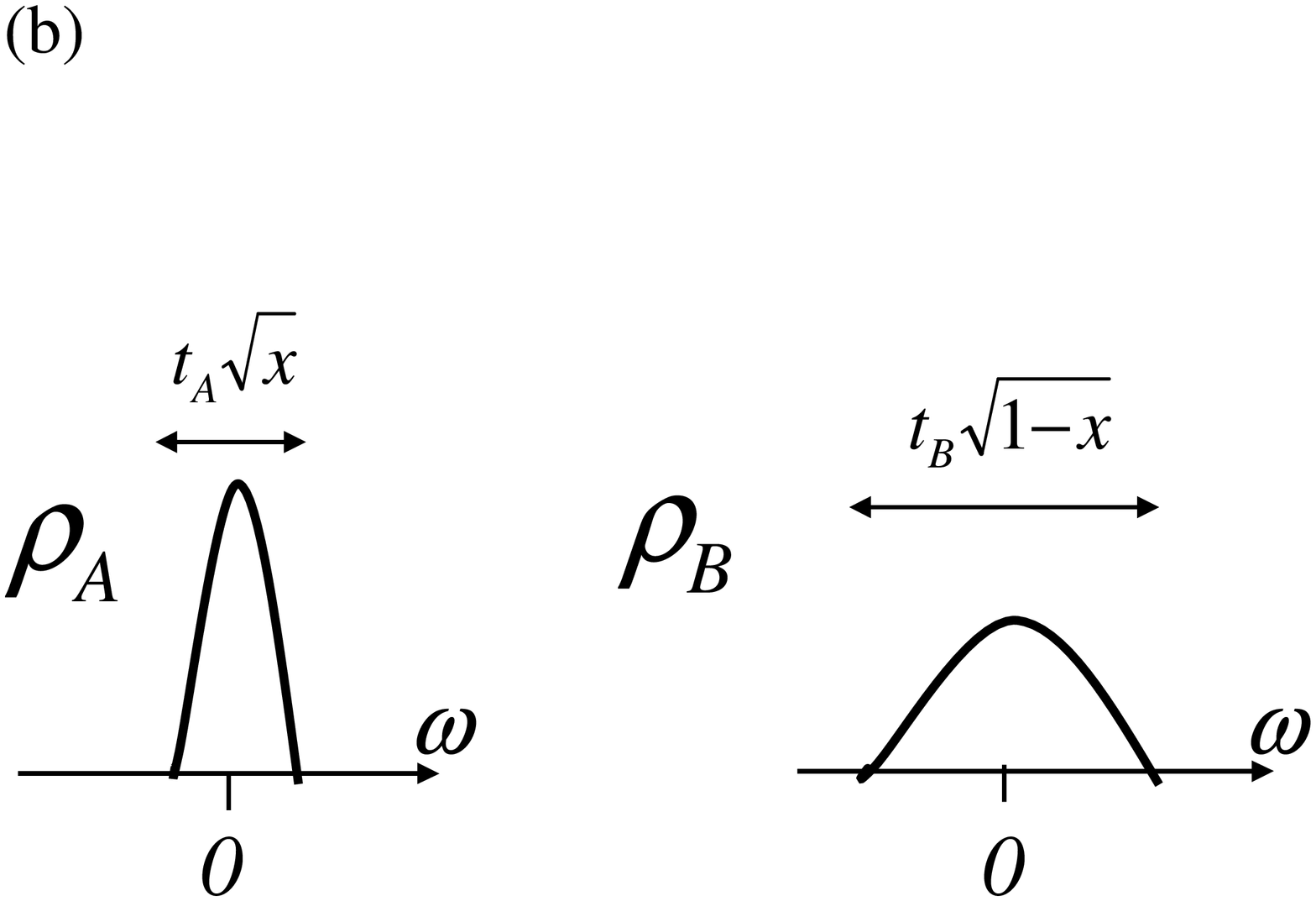}
\caption{Schematic plot of (a) the average local density of states 
of the conduction electrons $\rho=x\rho_{A}+(1-x)\rho_{B}$ 
in the limit $\alpha=t_{AB}/\sqrt{t_{A}t_{B}}>>1$, (b) the local densities 
of states of the conduction 
electrons $\rho_{A}$ (on a site $A$) and $\rho_{B}$ (on a site $B$) 
in the limit $\alpha=t_{AB}/\sqrt{t_{A}t_{B}}<<1$.
The concentration of atoms $A$ is assumed to be $x<1/2$. 
}
\label{FigureSchemasdostAB}
\end{figure}

{\it - Regime $\alpha>1$}\\
Integrating the local DOS, we define the average densities 
of electrons on sites $A$ and $B$ in the Fermi sea, as function of the 
Fermi energy $E_F$
\begin{eqnarray}
n_{A/B}
&\equiv& 
c_{A/B}\int_{-\infty}^{E_F}d\omega \rho_{A/B}(\omega)~, 
\end{eqnarray}
where $c_{A}=x$ and $c_{B}=1-x$. 
We analyse in FIG.~\ref{dosfreetab10.ps} the microscopical 
origin of the satellites and central peak 
in the averaged DOS for large values of $\alpha$. 
It appears that the two satellite peaks correspond to electronic excitations
which are equally distributed over $A$ and $B$ sites, while the central peak 
is due to excitations of electrons on the majority $B$ sites. 
Our interpretation is the following: when $\alpha>1$, the electronic 
states with the lowest energy are obtained by forming $A-B$ bonds. 
The deepest levels of the Fermi sea thus correspond to electronic wave 
functions localized on clusters formed by alternating $A$ and $B$ 
atoms. In the following, we call the latter "$AB$ clusters". 
We interprete the two satellite peaks as bonding and anti-bonding 
electronic states formed in these $AB$ clusters. 
In the large $z$ limit, each $A$ atom can be associated with a neighboring 
atom $B$. Choosing the latter to point into the same (arbitrary) "direction" 
garantees that 
the so formed $AB$ clusters contain exactely the same number of $A$ and $B$ atoms. 
As a consequence, the statistical weight of the $AB$ clusters is $2x$, which
is twice the concentration of the minority atoms $A$. 
Whether a given $B$ site belongs to an $AB$ cluster or to the embedding surface 
is left open. Nevertheless, the satellite peaks characterizing the DOS 
for $\alpha >>1$ have precisely the statistical weight $2x$, which is equaly
distributed over $A$ and $B$ atoms
(see FIG.~\ref{dosfreetab10.ps} and FIG.~\ref{FigureSchemasdostAB}~(a)). 
The remaining $B$ atoms constitute hypersurfaces embedding the $AB$ clusters. 
Considering that the dimension of the system is proportional to
$z$, the embedding surfaces provide in the large $z$ limit
a contribution to the DOS (central peak) which is
qualitatively similar to the one obtained for $\alpha=1$ 
(i.e., semi-elliptic here). 
The spectral weight of the latter is $1-2x$. 
Gaps in the local DOS, with two satellites and a central
peak appear only above a critical value of $\alpha$. This is seen in  
FIG.~\ref{FigurelocaldosrandomTB}. 

{\it - Regime $\alpha <1$}\\
In the limit $\alpha\to 0$, the two subsystems $A$ and $B$ decouple, and the 
averaged local DOS $\rho_{A}$ and $\rho_{B}$ can be deduced from the 
DOS of a pure system by rescaling the 
energy as $t_{A}\sqrt{x}$ and $t_{B}\sqrt{1-x}$ respectively 
(see FIG.~\ref{FigureSchemasdostAB}~(b)). 
As a consequence, when $t_{A}=t_{B}$, electrons with the lowest energy 
occupy predominantly majority atoms. 
When the density of conduction electrons is sufficiently large, the Fermi 
level moves into a region where electrons occupy both $A$ and $B$ sites. 
For that reason, the regime 
$\alpha<1$ is qualitatively not very different from the regime 
$\alpha=1$, except in the limit of a low density of conduction
electrons.

\subsection{Diagonal randomness: Kondo alloy \label{subsectiondiagonal}}
We consider next the transition between a diluted and a dense Kondo system. 
The model is defined by the KAM Eq.~(\ref{HamiltonianKAM}), with a periodic 
hopping element for the conduction electrons $t\equiv t_{A}=t_{B}=t_{AB}$. 
The limit $x\to 0$ corresponds to a single impurity Kondo model (SIKM), 
while $x=1$ describes a Kondo lattice model (KLM). 
These models have been extensively studied~[\onlinecite{KondoHewson}] by 
using various approximations. When we consider a paramagnetic
ground state, two energy scales are 
required~[\onlinecite{Anderson2scales1}-\onlinecite{Anderson2scales2}-\onlinecite{Anderson2scales4}] 
in order to describe the low temperature physical properties: 
$T_{K}$ characterizes the onset of the Kondo effect, and $T_{FL}$ characterizes
the formation of a coherent Fermi liquid ground state. 

On one side, 
the exact solution of the SIKM, based on the Bethe Ansatz~[\onlinecite{NAndrei1}-\onlinecite{NAndrei4}] 
proves that these two scales are identical in the dilute limit $x\to 0$. 
The low temperature physical properties of the SIKM are thus universal 
(i.e., independent from the lattice structure, electronic filling and Kondo
coupling) as soon as all the energy scales are rescaled by $T_{K}$. 

On the other side, for the KLM these two energy scales are different. 
In earlier works it had been suggested that for small conduction electron
densities $T_{FL}$ is strongly reduced, i.e., to 
$T_{FL}\to T_{K}^{2}/t$ 
because of conduction electron "exhaustion" when singlet states 
form~[\onlinecite{exhaustionearly1}-\onlinecite{exhaustionearly2}-\onlinecite{Nozieresnoexhaustion}]. 
However this has turned out not to be the case and 
$T_{FL}\approx T_{K}$~[\onlinecite{Nozieresnoexhaustion}-\onlinecite{KondolatticelargeN}]. 

The KAM which is studied here allows for describing the crossover between the dilute
regime (with $T_{FL}/T_{K}=1$) and the dense regime (with $T_{FL}/T_{K}$ 
depending on the electronic filling). 
From the static local magnetic susceptibility at zero temperature 
$\chi_{loc}(T=0)\equiv
\lim_{T\to 0}\int_{0}^{1/T}d\tau
\langle {\bf S}(\tau)\cdot  {\bf S}(0)\rangle$, we define an energy scale 
$T_{0}\equiv 1/4\chi_{loc}(T=0)$, which provides a reasonable estimate of 
$T_{FL}$~[\onlinecite{KondolatticelargeN}]. 

\begin{figure}[!ht]
\includegraphics[angle=270,width=0.5\columnwidth]{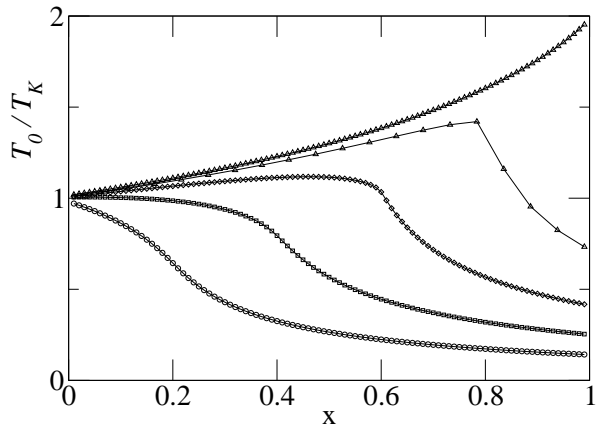}
\caption{Ratio between $T_0=1/4\chi_{loc}(T=0)\approx T_{FL}$ 
and $T_{K}$ as a function 
of the Kondo impurity concentration $x$. The different curves correspond to
different electronic filling: from top to bottom, $n_c=1; 0.8; 0.6; 0.4;
0.2$. The solid lines are guide for the eyes. 
The curves have been computed for hopping matrix elements 
$t\equiv t_{A}=t_{B}=t_{AB}$ and $J_K/t\sqrt{z}=1.5$.  
}
\label{FigureT0surTKnodisorder}
\end{figure}

FIG.~(\ref{FigureT0surTKnodisorder}) depicts the evolution of the ratio 
$T_{0}/T_{K}$ for different values of the electronic filling $n_c$. 
It starts from $T_{0}/T_{K}=1$ in the dilute limit $x\to 0$, 
as expected for the SIKM. 
With decreasing filling $n_c$ this ratio is lower 
in the dense limit $x=1$ corresponding to the KLM. 
This shows clearly that the crossover between the dense and the 
dilute regimes occurs when the concentration $x$ of Kondo impurities is equal 
to the electronic filling $n_{c}$. Note that similar results have been 
obtained recently by Kaul and Vojta for a $20\times 20$ sites square lattice 
(compare FIG.~(\ref{FigureT0surTKnodisorder}) with FIG.~(2) of 
Ref.~[\onlinecite{Vojta}]).

\begin{figure}[!ht]
\includegraphics[angle=0,width=0.5\columnwidth]{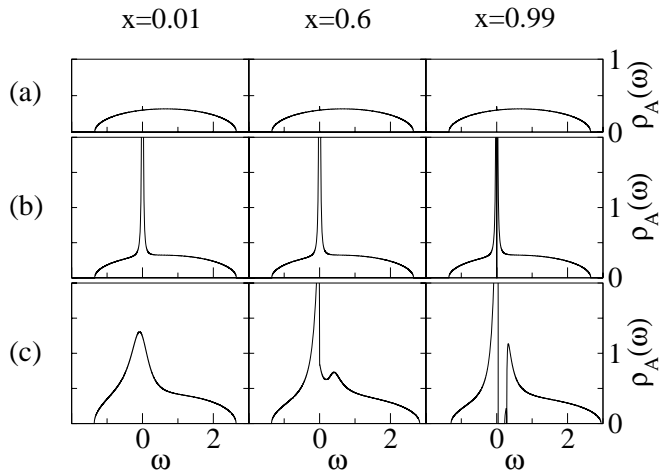}
\caption{Local electronic DOS on a magnetic site for an electronic 
filling $n_{c}=0.6$. Kondo impurity concentrations $x=0.01$ (diluted), 
$x=n_c=0.6$ (intermediate) and $x=0.99$ (dense). 
The energy unit is $\tilde{t}\equiv
t_{A}\sqrt{z}=t_{B}\sqrt{z}=t_{AB}\sqrt{z}$ 
and $J_K/\tilde{t}=1.5$.  
(a) Temperature $T>T_{K}$. 
(b) Temperature $T\approx T_{K}$. 
(c) Temperature $T=T_{K}/300$. 
}
\label{FigurerhoAnodisorder}
\end{figure}

The above feature can also be observed from the spectral function 
$\rho_{A}$, i.e., the local electronic DOS of a 
Kondo impurity. This is illustrated in FIG.~(\ref{FigurerhoAnodisorder}) 
for an electronic filling of $n_{c}=0.6$. 
Above the Kondo temperature, $\rho_{A}$ 
has the same semi-elliptic shape as in the absence of a localized spin. 
When $T\approx T_{K}$ a Kondo resonance develops 
at the Fermi level. Neither the quantitative value of $T_{K}$ 
nor the shape of the resonance depend on the 
concentration $x$. This suggests that the onset of the Kondo effect at 
$T_{K}$ is 
not a collective coherent effect, but results from incoherent scattering of 
conduction electrons by the magnetic impurities. 
When the temperature is decreased far below the Kondo temperature,  
a collective coherent screening takes place, which is accompanied by 
the onset of the Fermi liquid regime. 
For $T<<T_{K}$ the local DOS of $A$ sites depends on the concentration 
and shows in the dilute regime $x\to 0$ the standard Kondo resonance, 
while a gap occurs in the dense regime $x\approx 1$. 
We are aware that the presence of a Kondo gap is an
artefact of the mean-field approximation which we have introduced. Numerical 
studies of the KLM without this approximation show however, that a pseudo-gap will
form~[\onlinecite{KondolatticeCosti}]. 
Similarly to what we obtained for the ratio $T_{0}/T_{K}$, we find that the 
crossover between the dilute regime (without gap), and the dense regime 
(with a gap) occurs for $x=n_{c}$. 
FIG.~(\ref{FigurerhoAnodisorder}) depicts this behavior for an electronic
filling $n_{c}=0.6$. Similar results are obtained for 
$n_{c}=0.2; 0.4; 0.8$.

Even without the exhaustion argument, 
it appears that the crossover between a dilute and a
dense Kondo system occurs when the concentration $x$ of Kondo impurities is equal 
to the electronic filling $n_{c}$. 
We expect this result to apply also to the Anderson model, 
which still contains the charge degrees of freedom associated with the spins 
${\bf S}_{i}$ of the $A$ sites. 
This explains why the single impurity models, characterized by a unique low 
temperature scale $T_{FL}\approx T_{K}$ are still able to describe the low temperature 
properties of alloys with a significant concentration of magnetic 
impurities, here $A$ sites. 
As a consequence, in the experimental literature, the Kondo temperature of 
Kondo-like alloys is frequently estimated from different physical quantities, 
with no distinction beeing made 
between the quantities characterizing the Fermi liquid regime and 
those characterizing the onset of the Kondo effect. 
Nevertheless, two different energy scales, $T_{FL}$ and $T_{K}$, 
have been measured experimentaly and analysed for several rare earth alloys. 
For example, some very promising experimental results dedicated to an analysis 
of the crossover between diluted and dense impurity systems can be  
found in Ref.~[\onlinecite{ExperimentsYbLuAl}] (alloy 
$Yb_{1-x}Lu_{x}Al_{3}$), in Ref.~[\onlinecite{ExperimentsCeLaIrGe}] 
($Ce_{1-x}La_{x}Ir_{2}Ge_{2}$), or in Ref.~[\onlinecite{ExperimentsCeNiSi}] 
($CeNiSi_{2}$).

\subsection{Combined effects of randomness}
In this section we consider the KAM Eq.~(\ref{HamiltonianKAM}) with 
randomness of the electronic hopping matrix elements $t_{ij}$ and of 
Kondo alloying effects. 
We study the effect of alloying, described respectively
by the parameters $x$ and $n_{c}$, and of randomness, characterized by the
parameter $\alpha$ defined by Eq.~(\ref{Definitionalpha}). 
For the sake of clarity we discuss here only the behavior of the two 
low temperature energy scales introduced in the previous section: 
the Kondo temperature $T_{K}$,
characterizing the onset of incoherent singlet formation, and $T_{FL}$,
characterizing the onset of a coherent Fermi-liquid state. 
A good estimate of the latter can be obtained from the static 
local magnetic susceptibility at $T=0$ by the relation 
$T_{FL}\approx T_{0}=1/4\chi_{loc}(T=0)$. 

\subsubsection{Kondo temperature}
Next we study the effect of the concentration $x$ on the Kondo
temperature by considering various values of randomness $\alpha$ 
and electronic filling $n_{c}$. 
The main variations we observe for $T_{K}$ result from the effects of 
randomness on the non interacting local DOS $\rho_{A}$, which are 
analysed in Section \ref{subsectionoffdiagonal}. 

\begin{figure}[!ht]
\includegraphics[angle=270,width=0.5\columnwidth]{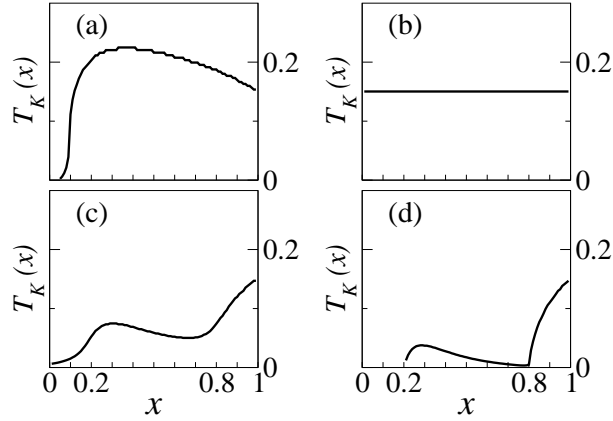}
\caption{Kondo temperature $T_{K}$ as a function of 
concentration $x$ of $A$ sites. Electronic filling is $n_{c}=0.4$. 
Kondo coupling is $J_{K}=1.5$. The energy scale is 
$\tilde{t}\equiv t_{A}\sqrt{z}=t_{B}\sqrt{z}$. 
(a) $\alpha=0$. (b) $\alpha=1$. (c) $\alpha=2$. (d) $\alpha=3$. 
}
\label{FigureTKdexdisorder}
\end{figure}

Figure~\ref{FigureTKdexdisorder} shows the dependence of the Kondo temperature
with respect to $x$. Without randomness (i.e., for $\alpha=1$), 
the Kondo temperature does not depend on $x$ 
(see FIG.~\ref{FigureTKdexdisorder}(b)). This reflects the fact that 
$T_{K}$ characterizes incoherent scattering of the conduction electrons  
on the spins ${\bf S}_{i}$. 
Due to the exponential dependence of $T_{K}$ on 
$\rho_{A}(E_{F})$ (see Eq.~(\ref{ExplicitTK})), 
with possible gap opening (in the regime 
$\alpha>1$) or bandwidth renormalisation (in the regime $\alpha<1$), 
$T_{K}$ can be strongly reduced in certain regimes of concentrations 
$x$ as soon as $\alpha\ne 1$. 

{\it - Regime $\alpha <1$}\\
A decrease of $\alpha$ does not really change the value of $T_{K}$, except 
in the dilute regime $x<<1$, where $T_{K}$ is strongly reduced 
(see FIG.~\ref{FigureTKdexdisorder}(a)). 
The reason is that when $x$ and $\alpha$ are small, the effective bandwidth of the 
local electronic DOS on $A$ sites is strongly reduced 
(see FIG.~\ref{FigureSchemasdostAB}~(b)).
Therefore the conduction electrons will fill the states at the non-magnetic
$B$ sites and $\rho_{A}(E_{F})$ is reduced. 

\begin{figure}[!ht]
\includegraphics[angle=270,width=0.5\columnwidth]{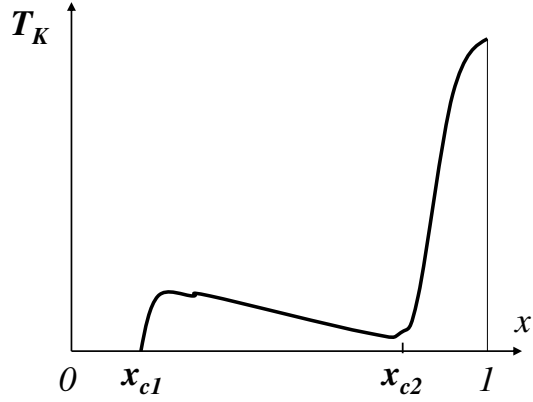}
\caption{Schematic plot of the Kondo temperature $T_{K}$ as a function of 
concentration $x$ of $A$ sites for strong $A-B$ electronic hopping 
($\alpha=t_{AB}/\sqrt{t_{A}t_{B}}>>1$).
Three regimes are recognized, separated by the concentrations $x_{c1}$ and 
$x_{c2}$. 
}
\label{FigureTKdexstrongdisorder}
\end{figure}

{\it - Regime $\alpha >1$}\\
When $\alpha$ increases, two critical concentrations $x_{c1}$ and $x_{c2}$ 
occur. They define three regimes (see FIG.~\ref{FigureTKdexdisorder} (c) and (d), 
and FIG.~\ref{FigureTKdexstrongdisorder}). 
This is a consequence of the complex structure of the local DOS discussed
in Section~\ref{subsectionoffdiagonal}, 
with a central and two satellite
peaks (see FIG.~\ref{FigureSchemasdostAB}~(a)). 

For $x>x_{c2}$, the Kondo or $A$ sites are in the majority. The Fermi level 
is in the central peak of the DOS,
which precisely corresponds to electronic excitations on $A$ sites. \\
When $x$ decreases from $1$ to $x_{c2}$, the DOS is modified such that the 
Fermi level approaches the band edge of the central peak. 
As a consequence, $T_{K}$ decreases and can even vanish for $x=x_{c}$ if 
$\alpha$ is large enough. 

For $x_{c1}<x<x_{c2}$, the Fermi level is positioned in a satellite peak, corresponding
to electronic excitations on both magnetic and non-magnetic sites.  
$T_{K}$ is finite but reduced compared to its value at $x=1$. 
For $x<x_{c1}$, the Fermi level is in the central peak, but the latter 
corresponds now to excitations on the non-magnetic $B$ sites which here are
in the majority. As a consequence, $\rho_{A}$ and $T_{K}$ are strongly reduced in
this regime and can even vanish if $\alpha$ is large enough. 

Relations between $x_{c1}$, $x_{c2}$ and the electronic
filling $n_{c}$ can be obtained from the observation that 
the spectral weight of the satellite peaks is proportional to the
concentration of the minority $A$ or $B$ sites. We thus find 
\begin{eqnarray}
x_{c1}&=& n_{c}/2 \\
x_{c2}&=& 1-n_{c}/2~. 
\label{Definitionsxc1xc2}
\end{eqnarray}
These relations corresponds to the critical values $x_{c1}=0.2$ and
$x_{c2}=0.8$ obtained numerically for $n_{c}=0.4$ 
(see FIGS.~\ref{FigureTKdexdisorder} (c) and (d)). They are also verified 
by numerical results for other electronic fillings. 

\subsubsection{Kondo versus Fermi liquid temperature}
\begin{figure}[!ht]
\includegraphics[angle=270,width=0.5\columnwidth]{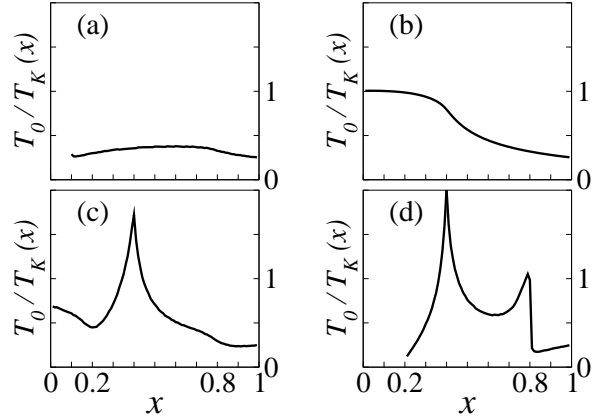}
\caption{
Ratio of $T_{0}\equiv 1/4\chi_{loc}(T=0)\approx T_{FL}$ and the Kondo temperature 
$T_{K}$ as a function of the 
concentration $x$ of $A$ sites. Electronic filling: $n_{c}=0.4$. 
Kondo coupling: $J_{K}=1.5$. The energy scale is 
$\tilde{t}\equiv t_{A}\sqrt{z}=t_{B}\sqrt{z}$. 
(a) $\alpha =0$. (b) $\alpha =1$. (c) $\alpha =2$. (d) $\alpha =3$. 
}
\label{FigureT0surTKdexdisorder}
\end{figure}

In the following we discuss the ratio between $T_{FL}$ and $T_{K}$. 
The effect of varying the electronic filling was discussed in 
Section~\ref{subsectiondiagonal} for $\alpha=1$. 
Figure~\ref{FigureT0surTKdexdisorder}(b) reproduces for convenience the plot of 
$T_{0}/T_{K}$ as a function of $x$ corresponding to $\alpha=1$ and $n_{c}=0.4$ 
shown in FIG.~\ref{FigureT0surTKnodisorder}. 
The other plots of FIG.~\ref{FigureT0surTKdexdisorder} depict the effect of
varying $\alpha$ with the electronic filling kept fixed. 
We obtained similar results with other values of $n_{c}$. For the sake of
clarity, we show here only the numerical results which 
we obtained for $n_{c}=0.4$. 

{\it - Regime $\alpha <1$}\\
The cross-over separating the dense and dilute regime is well
localized at $x=n_{c}$ and $\alpha=1$, and smoothed when $\alpha$ decreases 
(compeare FIG.~\ref{FigureT0surTKdexdisorder} (a) and (b)). 
At low concentrations, numerical calculations are less accurate. 
This is related to the strong decrease of $T_{K}$ when $x\to 0$ 
(see FIG.~\ref{FigureTKdexdisorder} (a)). For this reason, we are not able to 
provide reasonable results concerning $T_{0}/T_{K}$ for $x<0.1$. 

\begin{figure}[!ht]
\includegraphics[angle=270,width=0.5\columnwidth]{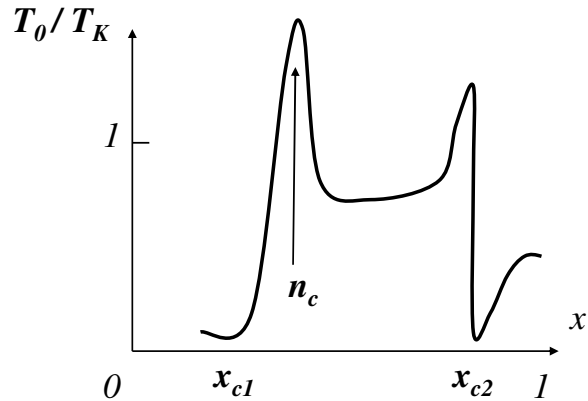}
\caption{Schematic plot of the ratio between 
$T_{0}\equiv 1/4\chi_{loc}(T=0)\approx T_{FL}$ and the Kondo temperature $T_{K}$ 
as a function of 
the Kondo impurity concentration $x$ for $\alpha >>1$ (i.e., large $A-B$ 
hopping matrix element). 
}
\label{FigureT0surTKdexstrongdisorder}
\end{figure}

{\it - Regime $\alpha >1$}\\
When $\alpha$ increases, two different effects occur: the crossover between the
dense and dilute regime becomes more pronounced around $x=n_{c}$ 
(see FIGS.~\ref{FigureT0surTKdexdisorder} (c) and (d)) and some 
"anomalies" occur at the critical concentrations $x_{c1}$ 
and $x_{c2}$ (see Eqs.~(\ref{Definitionsxc1xc2})). 
When $x$ is approaching any of these critical values, 
the ratio $T_{FL}/T_{K}$ becomes small (see $x=x_{c1}=0.2$ and 
$x=x_{c2}=0.8$ on FIG.~\ref{FigureT0surTKdexdisorder} (d)). 
Similar results were obtained for other values of $n_{c}$ 
($0.2; 0.4; 0.6; 0.8$). 
The general shape of the curve $T_{FL}/T_{K}$ in the regime $\alpha>1$ is 
depicted by FIG.~\ref{FigureT0surTKdexstrongdisorder}. 

When $T_{FL}/T_{K}\approx 1$, the physical properties of the system are
characterized by two universal temperature regimes: 
for $T>T_{K}$, the thermodynamic and transport properties are those of a 
light Fermi liquid (due to the conduction electrons), and the magnetic
properties are characteristical of free moments (for example, the 
magnetic susceptibility follows a Curie law). 
For $T<T_{FL}$, the physical properties correspond to an heavy Fermi
liquid. \\
When $T_{FL}/T_{K}<< 1$, an intermediate regime occurs, corresponding 
to $T_{FL}<T<T_{K}$, for which the temperature dependence of the 
physical properties might have non-Fermi-liquid or spin-liquid behavior. 

Note that the peak at $x=0.8$ in FIG.~\ref{FigureT0surTKdexdisorder} (d) 
(and more generaly at $x=x_{c2}$ on
FIG.~\ref{FigureT0surTKdexstrongdisorder}) 
has been explained in Ref~[\onlinecite{Kondopseudogap}] to which we refer.

\section{Summary and Conclusions}
The aim of this investigation has been to develop a description of Kondo
alloys, i.e., of systems with randomly placed Kondo ions of concentration
$x$. The values of $x$ were ranging from close to $x=0$ (diluted Kondo
impurities) to $x=1$ (Kondo lattice). Different hopping matrix elements were
assumed depending on wether the initial and final sites were Kondo sites or
nonmagnetic sites. By expressing the spin of a Kondo site in terms of
fermionic operators and by making a mean-field approximation we derived a
Hamiltonian which has in addition to the conduction band a narrow $f$ band
describing the low energy excitations of the system. The disorder of the
system was treated by a (dynamical) coherent potential approximation (CPA) and
by a dynamical mean-field theory (DMFT) approach. It was reconfirmed for the
special system considered here that both approaches yield identical
equations. For a practical application of the formalism a Bethe tree structure
was chosen for convenience. It corresponds to working with a semi-elliptic
density of states. Various quantities were calculated as function of the ratio 
$\alpha=t_{AB}/\sqrt{t_{A}t_{B}}$, i.e., of different degrees of off-diagonal
disorder and of Kondo ion concentration $x$. Among them were the local density
of states for the Kondo sites ($A$ atoms) and for the nonmagnetic sites ($B$
atoms). The averaged local densities of states depend strongly on the degree
of disorder and on concentration $x$. Of special interest was the case of
diagonal randomness only. In that case $t_{A}=t_{B}=t_{AB}$ and
$\alpha=1$. The ratio between the Fermi liquid temperature and the Kondo
temperature was investigated as function of $x$, i.e., ranging from the
impurity to the Kondo lattice limit. This ratio was shown to be of relevance
for a number of experiments. Finally, a detailed study was presented for the
case of a combined diagonal and off-diagonal disorder. In particular the
behavior of the Kondo temperature as function of $x$ was studied in
detail. The same holds true for the ratio of Fermi liquid to Kondo
temperature. The latter can have a strong monotonic behavior as function of
$x$ with peaks at different values of $x$. They depend on the particular
filling of the conduction band as well as on the parameter $\alpha$. Depending
on the ratio $T_{FL}/T_{K}$ we have obtained universal temperature regimes
where the system has either properties of a liquid with light fermions and
quasi-free local moments or of a liquid with heavy quasiparticles. There is
also a temperature regime possible with non-Fermi liquid behavior. A still
open question is under which conditions Luttinger's theorem is inapplicable
and how one can calculate in that case the volume enclosed by the Fermi
surface. As is well known that volume does not include the electrons which
form localized moments when we are in the regime of a light Fermi liquid. But
they must be included when we are in the regime of heavy
quasiparticles. Although the present investigation sheds some new light on
Kondo alloys and their prperties there are important issues remaining for the
future.

\acknowledgments

We would like to thank N.B. Perkins for stimulating discussions during 
the early phase of this work. 
We thank M. Vojta for comments which helped us to improve the 
manuscript. We thank M. Capone, K. Kikoin, M. Laad and C. Lacroix for 
fruitful discussions.

\appendix
\section{CPA equations \label{AppendixCPAequations}}
\subsection{Starting definitions}
Considering the mean-field Kondo Alloy Model defined by the 
effective Hamiltonian Eq.~(\ref{HamiltonianlargeN}), 
we define the Green's function, 
transfer integral, and local propagator matrices, as 
\begin{eqnarray}
{\widetilde{\bf G}}_{ij}= 
\left(
\begin{array}{ccc}
\hat{x}_{i}\hat{x}_{j}G_{ij}^{ff} & \hat{x}_{i}\hat{x}_{j}G_{ij}^{fc}& \hat{x}_{i}\hat{y}_{j}G_{ij}^{fc} \\
\hat{x}_{i}\hat{x}_{j}G_{ij}^{cf} & \hat{x}_{i}\hat{x}_{j}G_{ij}^{cc}& \hat{x}_{i}\hat{y}_{j}G_{ij}^{cc} \\
\hat{y}_{i}\hat{x}_{j}G_{ij}^{cf} & \hat{y}_{i}\hat{x}_{j}G_{ij}^{cc}& \hat{y}_{i}\hat{y}_{j}G_{ij}^{cc} 
\end{array}
\right)~,
&&
{\widetilde{\bf W}}=
\left(
\begin{array}{ccc}
0 & 0 & 0 \\
0 & t_{A} & t_{AB} \\
0 & t_{AB}& t_{B} 
\end{array}
\right)~,
\\
&&\nonumber \\
{\widetilde{\bf \Pi}}_{i}(i\omega)= 
\left(
\begin{array}{cc}
\hat{x}_{i} {\bf \Pi}_{A}(i\omega)
& 
\begin{array}{c}
0 \\ 
0
\end{array} \\
\begin{array}{lr}
0 & 0
\end{array} 
&
\hat{y}_{i}/(i\omega+\mu ) 
\end{array}
\right)~,
& {\rm with}~ &
{\bf \Pi}_{A}(i\omega) 
=
{\left(
\begin{array}{lr}
i\omega+\lambda & r\\
r & i\omega +\mu\\
\end{array}
\right)
}^{-1}~.
\end{eqnarray}
The projection operator $\hat{x}_{i}=1-\hat{y}_{i}$ is unity if 
$i$ refers to site $A$, and zero otherwise. 

\subsection{Equation of motion}
The equations of motion derived from the Hamiltonian 
Eq.~(\ref{HamiltonianlargeN}) for the scalar Green's functions $G_{ij}^{ff}$, 
$G_{ij}^{fc}$, and $G_{ij}^{cc}$, can be cast into the following 
matrix form: 
\begin{eqnarray}
{\widetilde{\bf G}}_{ij}(i\omega)
=
{\widetilde{\bf \Pi}}_{i}(i\omega)\delta_{ij}
+
{\widetilde{\bf \Pi}}_{i}(i\omega)~
{\widetilde{\bf W}}\gamma_{ij}~
{\widetilde{\bf \Pi}}_{j}(i\omega)
+
\sum_{l}
{\widetilde{\bf \Pi}}_{i}(i\omega)
{\widetilde{\bf W}}\gamma_{il}~
{\widetilde{\bf \Pi}}_{l}(i\omega)~
{\widetilde{\bf W}}\gamma_{lj}~
{\widetilde{\bf \Pi}}_{j}(i\omega)
+\cdots
\label{MotionequationCPA}
\end{eqnarray}
We assume that the Green's function and local propagator matrices can be 
inverted. This hypothesis can be satisfied by introducing a non zero parameter
$\delta$ such that $\hat{x}_{i}=1-\delta$ (or $x_i=\delta$) if $i$ is a site $A$ 
(respectively a site $B$). The limit $\delta=0$ is taken at the end of the 
calculations. 
Eq.~(\ref{MotionequationCPA}) can then be expressed as 
\begin{eqnarray}
{\left(
{\widetilde{\bf G}}^{-1}
\right)
}_{ij}(i\omega)
=
{\widetilde{\bf \Pi}}_{i}^{-1}(i\omega)\delta_{ij}
-{\widetilde{\bf W}}\gamma_{ij}~.
\label{MotionequationcompactCPA} 
\end{eqnarray}

\subsection{Local self-energy approximation}
Generalising the CPA procedure of Ref.~[\onlinecite{matrixCPABEB1}-\onlinecite{matrixCPABEB2}], 
we assume that the 
averaged Green's function matrix is characterised by a local 
$3\times 3$ self-energy matrix ${\widetilde{\bf \Sigma}}$, 
\begin{eqnarray}
{\left(
\left[
{\widetilde{\bf G}}(i\omega)
\right]^{-1}
\right)
}_{ij}
=
i\omega
{\widetilde{\bf I}}\delta_{ij}
-{\widetilde{\bf \Sigma}}(i\omega)\delta_{ij}
-{\widetilde{\bf W}}\gamma_{ij}~,  
\label{MotionequationcompactaverageCPA}
\end{eqnarray}
where ${\widetilde{\bf I}}$ is the $3\times 3$ identity matrix. 
Averaging over a random distribution of sites $A$ and $B$ restores the
translation symmetry of the underlying lattice. The averaged matrix Green's 
function is periodic in space, and its Fourier transform is 
\begin{eqnarray}
{\widetilde{\bf G}}_{{\bf k}}
\equiv 
\sum_{ij}e^{-i{\bf k}.({\bf R}_{i}-{\bf R}_{j})}
\left[ {\widetilde{\bf G}}_{ij} \right]~. 
\end{eqnarray}
Equation~(\ref{MotionequationcompactCPA}) can then be expressed as:  
\begin{eqnarray}
{\widetilde{\bf G}}_{{\bf k}}^{-1}(i\omega)
=
i\omega {\widetilde{\bf I}}
-
{\widetilde{\bf \Sigma}}(i\omega)
-
{\widetilde{\bf W}}\gamma_{{\bf k}}~. 
\label{GreenkannexeCPA}
\end{eqnarray}
Next we establish a complete set of self-consistent equations for 
the Green's function and the self-energy, 
whose matrix-elements are parametrised as follow
\begin{eqnarray}
{\widetilde{\bf \Sigma}}(i\omega)
\equiv
\left(
\begin{array}{cc}
{\bf \Sigma}_{A} (i\omega)
& 
\begin{array}{c}
\sigma_{1}(i\omega)\\ 
\sigma_{2}(i\omega)
\end{array} \\
\begin{array}{cc}
\sigma_{1}(i\omega) & \sigma_{2}(i\omega)
\end{array} 
&
\Sigma_{B}(i\omega)
\end{array}
\right)
\equiv
\left ( {
\begin{array}{ccc}
\Sigma_{A}^{ff}(i\omega) & \Sigma_{A}^{fc}(i\omega)& \sigma_{1}(i\omega) \\
\Sigma_{A}^{cf}(i\omega) & \Sigma_{A}^{cc}(i\omega)& \sigma_{2}(i\omega) \\
\sigma_{1}(i\omega) & \sigma_{2}(i\omega)& \Sigma_{B}^{cc}(i\omega)
\end{array}
}\right )~.
\end{eqnarray}

\subsection{Off-diagonal blocks of ${\widetilde{\bf \Sigma}}$: 
no double $A-B$ occupancy}
A first set of relations is obtained by expressing the local averaged 
Green's function in terms of their ${\bf k}-$dependent conterparts. 
\begin{eqnarray}
\left[
{\widetilde{\bf G}}_{00}(i\omega)
\right]
\equiv
\left ( {
\begin{array}{ccc}
xG_{A}^{ff}(i\omega) & xG_{A}^{fc}(i\omega) & 0 \\
xG_{A}^{cf}(i\omega) & xG_{A}^{cc}(i\omega) & 0 \\
0               & 0               & (1-x)G_{B}^{cc}(i\omega)
\end{array}
}\right )
=
\sum_{{\bf k}}
{\widetilde{\bf G}}_{{\bf k}}(i\omega)~. 
\end{eqnarray}
Here ${\widetilde{\bf G}}_{{\bf k}}$ is of the form of 
Eq.~(\ref{GreenkannexeCPA}). 
Even if our approach artificially introduces two conducting bands $A$ and $B$, 
the vanishing of the off-diagonal mixed $A-B$ matrix elements in 
$\langle {\widetilde{\bf G}}_{00}\rangle_{r}$ prevents a double association of 
$A$ and $B$ atoms to the same site. The two corresponding equations determine the 
self-energy off-diagonal blocks $\sigma_{1}$ and $\sigma_{2}$.

\subsection{Diagonal blocks of ${\widetilde{\bf \Sigma}}$: 
scattering of the effective medium}
Self-consistent relations for ${\bf \Sigma}_{A}$ 
and $\Sigma_{B}$ are 
obtained by requiring that the scattering of electrons of the effective 
medium by a given site vanishes on average. 
For a given random configuration of the alloy, and using 
Eqs.~(\ref{MotionequationcompactCPA}, 
\ref{MotionequationcompactaverageCPA}), 
the Green's function 
${\widetilde{\bf G}}$, is related to its average by 
\begin{eqnarray}
{\left(
{\widetilde{\bf G}}^{-1}
\right)
}_{ij}
=
{\left(
\left[
{\widetilde{\bf G}}
\right]^{-1}
\right)
}_{ij}
-
{\widetilde{\bf v}}_{i}\delta_{ij}~,
\end{eqnarray}
where 
\begin{eqnarray}
{\widetilde{\bf v}}_{i}(i\omega)
&=&
i\omega{\widetilde{\bf I}}
-{\widetilde{\bf \Pi}}_{i}^{-1}(i\omega)
-{\widetilde{\bf \Sigma}}(i\omega)
\nonumber \\
&& \\
&=&
\left ( {
\begin{array}{lcr}
i\omega -\Sigma_{A}^{ff}(i\omega) -(i\omega+\lambda)/\hat{x}_{i} &
 -\Sigma_{A}^{fc}(i\omega) - r/\hat{x}_{i} &
-\sigma_{1}(i\omega) \\
 -\Sigma_{A}^{cf}(i\omega) - r/\hat{x}_{i} &
i\omega -\Sigma_{A}^{cc}(i\omega) -(i\omega+\mu)/\hat{x}_{i} &
-\sigma_{2}(i\omega) \\
-\sigma_{1}(i\omega) &
-\sigma_{2}(i\omega) &
i\omega -\Sigma_{B}^{cc}(i\omega) -(i\omega+\mu)/\hat{y}_{i} 
\end{array}
}\right )~.
\nonumber
\end{eqnarray}
The single-site scattering $T-$matrix on site $i$ is given by
\begin{eqnarray}
{\widetilde{\bf \tau}}_{i}
&=&
{\left(
{\widetilde{\bf I}}
-
{\widetilde{\bf v}}_{i}
\left[
{\widetilde{\bf G}}_{ii}
\right]
\right) 
}^{-1}
{\widetilde{\bf v}}_{i}\\
&=&
{\left( 
{\widetilde{\bf v}}_{i}^{-1}
-
\left[
{\widetilde{\bf G}}_{ii}
\right]
\right)
}^{-1}~. 
\end{eqnarray}
As mentioned above, the required matrix-inversions can 
be performed by introducing a non-zero parameter $\delta$ such that 
$\hat{x}_{i}$ is equal to $1-\delta$ (site $A$) or $\delta$ (site $B$). 
We now consider the limit $\delta=0$ and we express the scattering 
matrix ${\widetilde{\bf \tau}}_{A}$ and 
${\widetilde{\bf \tau}}_{B}$ on a site $A$ 
(respectively $B$). 
We find 
\begin{eqnarray}
{\widetilde{\bf \tau}}_{A}
= 
\left(
\begin{array}{cc}
{\left (
{\bf v}_{A}^{-1} -x{\bf G}_{A}
\right )
}^{-1}
& 
\begin{array}{c}
0 \\ 
0
\end{array} \\
\begin{array}{lr}
0 & 0
\end{array} 
&
-\left(
(1-x)G_{B}^{cc}
\right)^{-1}
\end{array}
\right)~,
\end{eqnarray}
and 
\begin{eqnarray}
{\widetilde{\bf \tau}}_{B}
= 
\left(
\begin{array}{cc}
-
{\left(
x{\bf G}_{A}
\right )
}^{-1}
& 
\begin{array}{c}
0 \\ 
0
\end{array} \\
\begin{array}{cc}
0 & 0
\end{array} 
&
\left(
v_{B}^{-1}-(1-x)G_{B}^{cc}
\right)^{-1}
\end{array}
\right)~, 
\end{eqnarray}
with 
\begin{eqnarray}
&{\bf v}_{A}(i\omega)
&=
-
\left ( {
\begin{array}{lr}
\lambda & 
r \\
r & 
\mu
\end{array}
} \right )
-{\bf \Sigma}_{A}(i\omega)
~, 
\\
&&\nonumber \\
&{\bf G}_{A}(i\omega)
&=
\left ( {
\begin{array}{lr}
G_{A}^{ff}(i\omega)& 
G_{A}^{fc}(i\omega)\\
G_{A}^{cf}(i\omega)& 
G_{A}^{cc}(i\omega)
\end{array}
} \right )~, 
\\
&&\nonumber \\
&v_{B}(i\omega)
&=
-\mu - \Sigma_{B}(i\omega)~. 
\end{eqnarray}
We obtain the CPA equations for ${\bf \Sigma}_{A}$ and $\Sigma_{B}$ by setting
$
\left[
{\widetilde{\bf \tau}}_{i}
\right]
=
x{\widetilde{\bf \tau}}_{A}
+
(1-x){\widetilde{\bf \tau}}_{B}
=0
$. The result is 
\begin{eqnarray}
{\bf \Sigma}_{A}(i\omega)
&=&
-
\left ( {
\begin{array}{lr}
\lambda & r \\
r & \mu 
\end{array}
}\right )
-{{(1-x)}\over{x}}
{\bf G}_{A}^{-1}(i\omega)~, 
\nonumber\\
&&\\
\Sigma_{B}(i\omega)
&=&
-\mu 
-{x\over{(1-x)G_{B}^{cc}(i\omega)}}~. 
\nonumber
\end{eqnarray}



\begin{thebibliography}{99}

\bibitem{reviewHF1} 
P. Fulde, P. Thalmeier, and G. Zwicknagl, {\it Strongly correlated electrons}, 
Solid State Physics {\bf 60} (Elsevier, 2006). 
\bibitem{reviewHF2} 
G. R. Stewart, Rev. Mod. Phys. {\bf 73}, 797 (2001). 
\bibitem{KondoHewson} 
A.C. Hewson, {\it The Kondo Problem to Heavy Fermions} 
(Cambridge University Press, Cambridge, England, 1993). 
\bibitem{SchriefferWolf} 
J.R. Schrieffer, and P.A. Wolf, Phys. Rev. {\bf 149}, 491 (1966). 
\bibitem{Kondoanddisorder1} 
V. Dobrosavljevic, T.R. Kirkpatrick, and G. Kotliar, 
Phys. Rev. Lett. {\bf 69}, 1113 (1992).  
\bibitem{Kondoanddisorder2} 
E. Miranda and V. Dobrosavljevic, Phys. Rev. Lett. {\bf 78}, 290 (1997). 
\bibitem{Kondoanddisorder5} 
S. Kettemann, and E.R. Mucciolo, JETP Letters {\bf 83}, 240 (2006). 
\bibitem{Kondoanddisorder2bis} 
A.M. Sengupta, and A. Georges,  Phys. Rev. B {\bf 52}, 10295 (1995). 
\bibitem{Kondoanddisorder3} 
A.H. Castro Neto, G. Castilla, and B.A. Jones, 
Phys. Rev. Lett. {\bf 81}, 3531 (1998). 
\bibitem{Kondoanddisorder4} 
S. Burdin, D.R. Grempel, and A. Georges, Phys. Rev. B {\bf 66}, 045111
(2002). 
\bibitem{Vojta} 
R.K. Kaul, and M. Vojta. cond-mat/0603002. 
\bibitem{matrixCPABEB1} 
J.A. Blackman, D.M. Esterling, and N.F. Berk, Phys. Rev. B {\bf 4}, 2412
(1971). 
\bibitem{matrixCPABEB2} 
D.M. Esterling,  Phys. Rev. B {\bf 12}, 1596 (1975). 
\bibitem{reviewDMFT1} 
A. George, G. Kotliar, W. Krauth, and M.J. Rozenberg, 
Rev. Mod. Phys. {\bf 68}, 13 (1996). 
\bibitem{reviewDMFT2} 
W. Metzner and D. Vollhardt, Phys. Rev. Lett. {\bf 62}, 324 (1989).  
\bibitem{Kondomeanfield1} 
C. Lacroix and M. Cyrot, Phys. Rev. B {\bf 20}, 1969 (1979).  
\bibitem{Kondomeanfield2} 
B. Coqblin, M.A. Gusmao, J.R. Iglesias, C. Lacroix, A. Ruppenthal,
A.S.D. Simoes, Physica (Amsterdam) {\bf 282B}, 50 (2000). 
\bibitem{KondolargeN1} 
N. Read, D.M. Newns, and S. Doniach, Phys. Rev. B {\bf 30}, 3841 (1984). 
\bibitem{KondolargeN5} 
D.M. Newns and N. Read, Adv. Phys. {\bf 36}, 799 (1987). 
\bibitem{KondolargeN6} 
P. Coleman, Phys. Rev. B {\bf 35}, 5072 (1987). 
\bibitem{scalarCPA1} 
See, for example, P. Fulde, {\it Electron correlations in molecules and
  solids}, 3rd enlarged ed. (Springer series in Solid State Sciences, Berlin, 1995). 
\bibitem{scalarCPA2} 
P. Soven, Phys. Rev. {\bf 156}, 809 (1967). 
\bibitem{scalarCPA3} 
D.W. Taylor, Phys. Rev. {\bf 156}, 1057 (1967). 
\bibitem{scalarCPA4} 
B. Velicky, S. Kirkpatrick, and H. Ehrenreich, 
Phys. Rev. {\bf 175}, 747 (1968). 
\bibitem{scalarCPA5} 
H. Hasegawa, and J. Kanamori, J. Phys. Soc. Jpn. {\bf 31}, 382 (1971). 
\bibitem{KondolatticelargeN} S. Burdin, A. George, and D.R. Grempel, 
Phys. Rev. Lett. {\bf 85}, 1048 (2000).  
\bibitem{ClusterDMFTCPA}
M. Jarrell, and H.R. Krishnamurthy, Phys. Rev. B {\bf 63}, 125102 (2001). 
\bibitem{Kakehashi} 
Y. Kakehashi, Phys. Rev. B {\bf 66}, 104428 (2002). 
\bibitem{DasSarma} 
A. Chattopadhyay, S. Das Sarma, and A.J. Millis, Phys. Rev. Lett. {\bf 87},
227202 (2001). 
\bibitem{Anderson2scales1} 
M. Jarrell, H. Akhlaghpour, and T. Pruschke, Phys. Rev. Lett. {\bf 70}, 1670 (1993). 
\bibitem{Anderson2scales2} 
A.N. Tahvildar-Zadeh, M. Jarrell, and J.K. Freericks, Phys. Rev. Lett. {\bf 80}, 5168 (1998). 
\bibitem{Anderson2scales4} 
Th. Pruschke, R. Bulla, and M. Jarrell, Phys. Rev. B {\bf 61}, 12799 (2000). 
\bibitem{NAndrei1} 
N. Andrei, Phys. Rev. Lett. {\bf 45}, 379 (1979). 
\bibitem{NAndrei4} 
P. Wiegman, JETP Letter {\bf 31}, 392 (1980). 
\bibitem{exhaustionearly1} 
P. Nozi\`eres, Ann. Phys. (Paris) {\bf 10}, 19 (1985). 
\bibitem{exhaustionearly2} 
P. Nozi\`eres, Eur. Phys. B {\bf 6}, 447 (1998). 
\bibitem{Nozieresnoexhaustion} 
P. Nozi\`eres, J. Phys. Soc. Jpn. {\bf 74}, 4 (2005). 
\bibitem{KondolatticeCosti} 
T. Costi, and N. Manini, J. Low Temp. Phys. {\bf 126}, 835 (2002). 
\bibitem{ExperimentsYbLuAl} 
E.D. Bauer, C.H. Booth, J.M. Lawrence, M.F. Hundley, J.L. Sarrao, 
J.D. Thompson, P.S. Riseborough, T. Ebihara, 
Phys. Rev. B {\bf 69}, 125102 (2004).
\bibitem{ExperimentsCeLaIrGe} 
R. Mallik, E.V. Sampathkumaran, P.L. Paulose, J. Dumschat, and G. Wortmann, 
Phys. Rev. B {\bf 55}, 3627 (1997). 
\bibitem{ExperimentsCeNiSi} 
E.D. Mun, Y.S. Kwon, and M.H. Jung, Phys. Rev. B {\bf 67}, 033103 (2003).
\bibitem{Kondopseudogap} 
S. Burdin, and V. Zlatic, cond-mat/0212222. 


\end{thebibliography}
\end{document}